\documentclass[aps,prx,reprint, superscriptaddress]{revtex4-2}
\usepackage{bbold}
\usepackage{comment}
\usepackage[colorlinks, citecolor=red]{hyperref}
\usepackage{hyperref}
\usepackage[utf8]{inputenc}
\usepackage[T1]{fontenc}    %
\usepackage[english]{babel}
\usepackage[pdftex]{graphicx}
\usepackage{amsmath,amssymb,amsthm,bbm, mathtools} %
\usepackage{mathrsfs}
\usepackage{xcolor}
\usepackage[normalem]{ulem}

\begin{document}

\title{High-fidelity spin qubit shuttling via large spin-orbit interaction}

\author{Stefano Bosco}
\email{stefano.bosco@unibas.ch}
\author{Ji Zou}
\author{Daniel Loss}
\affiliation{Department of Physics, University of Basel, 4056 Basel, Switzerland}

\begin{abstract}
Shuttling spins with high fidelity is a key requirement to scale up semiconducting quantum computers, enabling qubit entanglement over large distances and favouring the integration of control electronics on-chip. To decouple the spin from the unavoidable charge noise, state-of-the-art spin shuttlers try to minimize the inhomogeneity of the Zeeman field. 
However, this decoupling is challenging in otherwise promising quantum computing platforms such as hole spin qubits in silicon and germanium, characterized by a large spin-orbit interaction and electrically-tunable qubit frequency.
In this work, we show that, surprisingly, the large inhomogeneity of the Zeeman field stabilizes the coherence of a moving spin state, thus enabling high-fidelity shuttling also in these systems. 
We relate this enhancement in fidelity to the deterministic dynamics of the spin which filters out the dominant low-frequency contributions of the charge noise. 
By simulating several different scenarios and noise sources, we show that this is a robust phenomenon generally occurring at large field inhomogeneity. 
By appropriately adjusting the motion of the quantum dot, we also design realistic protocols enabling faster and more coherent spin shuttling. 
Our findings are generally applicable to a wide range of setups and could pave the way toward large-scale quantum processors.
\end{abstract}

\maketitle

\section{Introduction}
Spin qubits confined in silicon and germanium quantum dots are front-runners in the 
race toward large-scale quantum computers~\cite{Philips2022,Noiri2022,Takeda2022,Xue2022,ZajacResonantlydrivenCNOT2018,hendrickx2020four,hendrickx2020fast,watzinger2018germanium,Jirovec2021,liles2023singlet,doi:10.1126/sciadv.abn5130,doi:10.1146/annurev-conmatphys-030212-184248,RevModPhys.95.025003,scappucci2020germanium,geyer2022two}. Their demonstrated compatibility with industry-level CMOS processing~\cite{geyer2022two,Zwerver2022,maurand2016cmos,camenzind2021spin} and their high-temperature operations~\cite{camenzind2021spin,PhysRevLett.121.076801,Yang2020,Petit2020,undseth2023hotter} make these systems ideal for scalability and co-integration with control electronics~\cite{Vandersypen2017,Xue2021,PhysRevApplied.18.024053,Bavdaz2022,Gonzalez-Zalba2021,kunne2023spinbus}.
The small footprint of spin qubits, typically a few tens of nanometers, however, imposes demanding technological constraints for the classical hardware and requires dense multi-layered architectures that add significant extra complexity to the process~\cite{Veldhorst2017,doi:10.1126/sciadv.aar3960,Borsoi2023,john2023bichromatic}.

These constraints are significantly relaxed by introducing quantum links coupling distant spins that are placed micrometers apart~\cite{Vandersypen2017}. This long-range connectivity can be achieved in various ways, including for example virtual couplings enabled by photons in superconducting cavities~\cite{mi2018coherent,Landig2018,doi:10.1126/science.aaa3786,yu2022strong,PhysRevX.12.021026,PhysRevLett.130.137001,PhysRevLett.108.190506,PhysRevLett.129.066801,PhysRevLett.118.147701,PhysRevB.100.161303,PhysRevB.100.081412,PhysRevB.96.235434,DRkloeffel2,PhysRevResearch.3.013194,Cubaynes2019,Borjans2020}, Luttinger liquids~\cite{doi:10.1063/1.4868868,PhysRevLett.93.126804,PhysRevB.100.035416,PhysRevB.96.115407,PhysRevApplied.12.014030,PhysRevB.93.075301}, floating gates~\cite{PhysRevX.2.011006,PhysRevB.95.245422}, and magnetic systems~\cite{PhysRevX.3.041023,PRXQuantum.2.040314,PhysRevB.99.140403,PhysRevB.105.245310,PhysRevResearch.5.033166,PhysRevB.106.235409}. Correlated dissipative coupling emerging by appropriately engineering the spin coupling to a bosonic bath has also been proposed as a viable route to entangle distant qubits~\cite{PhysRevB.106.L180406,zou2023spatially}.  All these possibilities, however, require external components that are not straightforward to integrate into the conventional CMOS processes, therefore hindering the competitive advantage of spin qubits. 

\begin{figure}[t]
\centering
\includegraphics[width=0.49\textwidth]{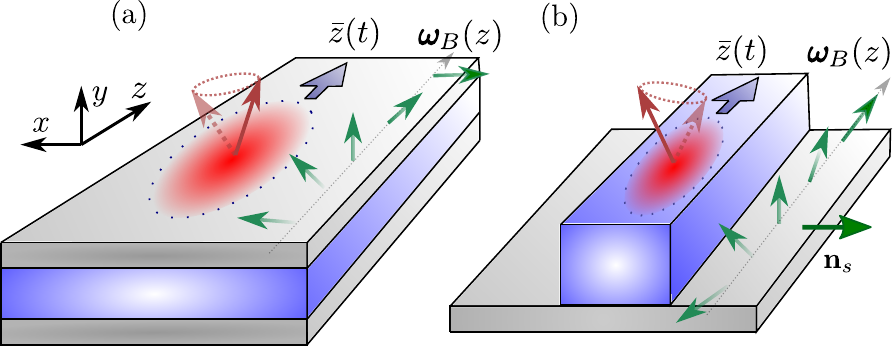}
\caption{
Sketch of our moving spin qubits. A particle confined in a quantum dot centered at the time-dependent position $\bar{z}(t)$ is shuttled along the $z$-direction in a planar germanium heterostructure (a) or in a silicon fin field effect transistor (b). During the motion, the spin of the particle (red arrows) precesses because of an inhomogeneous Zeeman field $\pmb{\omega}_B(z)$ (green arrows), which can be caused by a space-dependent $g$-tensor and magnetic field (a) or by a large spin-orbit interaction (SOI) with direction $\textbf{n}_{s}$ (b).
The fast SOI-induced dynamics of the spin filters out dangerous low-frequency noise and substantially boosts the fidelity of the spin shuttling.
\label{fig:sketch}
}
\end{figure}

On the other hand, shuttling spins across the chip provides a viable and CMOS-compatible way to link qubits in a sparse array~\cite{PRXQuantum.4.030303,Yoneda2021,Mills2019,Fujita2017,van2023coherent,Seidler2022,struck2023spin,PRXQuantum.4.020305}. The fidelity of this operation is determined by the noise that the spin experiences during shuttling and, in current devices, this noise is predominantly related to random fluctuations of the electrostatic environment. Because spin-orbit interactions (SOI) directly couple the spin degree of freedom to these charge fluctuations, current experiments try to minimize the SOI to maximize the shuttling fidelity. This approach is challenging in hole-based spin qubits, whose predominant feature is their large SOI~\cite{DRkloeffel1,DRkloeffel3,Froning2021,Wang2022,PhysRevResearch.3.013081,Liu2023,PhysRevB.103.125201,PhysRevB.104.115425,PhysRevB.105.075308,PhysRevB.105.075308,PhysRevB.106.235408,PhysRevLett.131.097002,PhysRevB.106.235426,bosco2023phase,sarkar2023electrical}. The tunability of their SOI enables sweet spots~\cite{PRXQuantum.2.010348,PhysRevLett.127.190501,Piot2022,PhysRevApplied.18.044038,wang2022modelling,michal2022tunable,Wang2021,hendrickx2023sweet} where the SOI can be turned off, however, for shuttling operations this optimization requires a demanding fine-tuning of the electrostatic potential over wide areas.  

In this work, we show that, surprisingly, large SOI and inhomogeneity of the Zeeman field can substantially \textit{enhance} the shuttling fidelity. This improvement depends on the coherent dynamics imprinted by the SOI on the spin state. The spin moving in a large SOI field rotates quickly in a deterministic and controllable way and this motion provides an intrinsic dynamical decoupling, filtering out the dominant, low-frequency contribution of the noise, thus boosting shuttling fidelity. The high spin shuttling fidelities reached in our shuttling scheme are qualitatively independent of the type and spatial distribution of the noise sources, and can be reached also by moving spin in an inhomogeneous Zeeman field for example produced by varying $g$-tensors~\cite{PhysRevLett.120.137702,PhysRevB.104.235303} or micromagnets~\cite{Yoneda2021, Philips2022, 10.1063/5.0139670}, opening up to effective SOI-driven improvements of shuttling fidelity in electron spin qubits in silicon and germanium.
Expanding on these ideas, we propose optimal protocols to leverage the SOI to further dynamically decouple the moving spin from the environment,  rendering the shuttling faster and at the same time more coherent, and paving the way towards high-fidelity shuttling of hole spin qubits for the large-scale quantum processors.

This manuscript is organized as follows. 
In Sec.~\ref{sec:Model}, we introduce our general  model describing spins shuttling in  inhomogeneous SOI and Zeeman fields. Our theory captures the spin dynamics in a wide variety of setups, including the silicon and germanium spin qubits in fin field-effect transistors and heterostructures sketched in Fig.~\ref{fig:sketch}. 
In Sec.~\ref{sec:Zeeman}, we specialize our discussion on inhomogeneous Zeeman fields only. This simple case provides a valuable intuitive understanding of the coherent and incoherent time evolution of the spin, and of the effect of different sources of noise. We expand the discussion in  Sec.~\ref{sec:SOI}, by including an inhomogeneous SOI field, nicely describing realistic hole-based silicon and germanium devices. We show that large effective SOI is beneficial to reduce the effect of noise during shuttling and, as proposed in Sec.~\ref{sec:Dyn}, it can be further leveraged in alternative shuttling schemes that dynamically decouple the spin from low-frequency noise. These protocols enable a faster motion of the spins and substantially boost the shuttling coherence of a wide range of materials and systems presenting large inhomogeneities of Zeeman fields.

\section{\label{sec:Model} Theoretical model}
In this work, we analyze spin qubits confined in moving quantum dots, as sketched in Fig.~\ref{fig:sketch}. 
The dynamic of the spin along the direction of motion ($z$-direction) is accurately modeled by the effective one-dimensional Hamiltonian
\begin{equation}
\label{eq:H_1D_general}
    H_\text{1D}=\frac{p^2}{2m}+\frac{m\omega_o^2}{2}[z-\bar{z}(t)]^2- \{\textbf{v}(z), p\} \cdot\pmb{\sigma}+ \frac{\hbar \tilde{\pmb{\omega}}_B(z)\cdot\pmb{\sigma}}{2} \ .
\end{equation}

This Hamiltonian  describes a quantum dot with harmonic frequency $\omega_o$ and width $l=\sqrt{\hbar/m\omega_o}$, whose center of mass $\bar{z}(t)$ is shifted  time-dependently. 
This moving electric potential is experimentally implemented in conveyer-mode shuttling architectures~\cite{Seidler2022,struck2023spin,PRXQuantum.4.020305}. In this work we restrict ourselves to this type of shuttling, however, we expect that our results can be generalized also to buckled-brigade shuttling~\cite{PRXQuantum.4.030303,Yoneda2021,Mills2019,Fujita2017,van2023coherent,PhysRevB.104.075439}.
During its motion, the spin experiences inhomogeneous spin-orbit and Zeeman fields, described by the vectors of spin-orbit velocities $\textbf{v}(z)$ and Larmor frequencies $\tilde{\pmb{\omega}}_{B}(z)$, respectively. We anticipate that the local Zeeman field of the nanostructure $\tilde{\pmb{\omega}}_{B}(z)$ differs from the local qubit splitting $\pmb{\omega}_{B}(z)$ by a correction arising from the confinement in the $z$-direction~\cite{PhysRevResearch.3.013081,PhysRevB.104.115425,PhysRevB.105.075308,PhysRevB.106.235408}, see Eq.~\eqref{eq:H_IZ}.  Here,  $m$ is the effective mass along $z$, $p=-i\hbar\partial_z$ is the momentum operator in the direction of motion, $\pmb{\sigma}=(\sigma_1,\sigma_2,\sigma_3)$ is the vector of Pauli matrices, and $\{ a, b\}=(a b+ba)/2$ is the symmetrized anticommutator that guarantees the hermiticity of $H_\text{1D}$. 

Our Eq.~\eqref{eq:H_1D_general} generally captures the response of a wide variety of different setups, including the ones in Fig.~\ref{fig:sketch}.
In particular, in this work, we focus primarily on hole spin qubit architectures, where the effective parameters originate from the mixture of heavy and light holes in the valence band caused by kinetic energy and strain and described by the Luttinger-Kohn and Bir-Pikus Hamiltonian.  
In one-dimensional hole channels in silicon and germanium, the SOI velocity $\textbf{v}(z)$ is large, yielding experimentally measured SOI lengths $\lambda_s= \hbar/m|\textbf{v}|$ of tens of nanometers, comparable with the quantum dot width $l$~\cite{Froning2021, Wang2022,PhysRevResearch.3.013081,Liu2023,geyer2022two,camenzind2021spin}. In planar hole nanostructures, the SOI is generally smaller although it can be enhanced by device engineering~\cite{PhysRevB.104.115425,PhysRevB.105.075308,PhysRevB.106.235408,PhysRevB.107.L161406,PhysRevB.106.235426,PhysRevB.103.085309,PhysRevLett.131.097002,rodriguez2023linear,Assali}. However, in these systems, the effective Zeeman field $\tilde{\pmb{\omega}}_B^h(z)=\mu_B\hat{g}(z)\textbf{B}/\hbar$ is also largely inhomogeneous because of the space-dependent and electrically tunable $g$ tensor $\hat{g}(z)$, which rotates the energetically preferred quantization axes at different location also when the externally applied magnetic field $\textbf{B}$ is homogeneous~\cite{PhysRevB.104.235303,van2023coherent,hendrickx2023sweet,Jirovec2021,PhysRevLett.120.137702,liles2023singlet}. 
We stress that our model also directly describes  electron spin qubits moving in an inhomogeneous magnetic field provided, for example, by micromagnets~\cite{Philips2022,Noiri2022,Takeda2022,Xue2022,ZajacResonantlydrivenCNOT2018}. In this case, similarly to planar hole heterostructures, the SOI $\textbf{v}(z)$ is small, and the leading contribution to the spin dynamics is the inhomogeneous Zeeman field $\tilde{\pmb{\omega}}_B^e(z)=\mu_B g \textbf{B}(z)/\hbar$.

Throughout this work, we restrict ourselves to adiabatically moving quantum dots, and consider shuttling velocities that are slow compared to the orbital energy gap. The small corrections to our model arising from non-adiabaticity in the orbital degrees of freedom and an exact solution of a simple case where this condition is lifted are discussed in detail in Appendix~\ref{App:Non_adiabatic}.
We note that for holes this condition is $\hbar\partial_t \bar{z}/l\ll \hbar\omega_o\sim 1$~meV, while for electrons in silicon and germanium this condition is more stringent and we require $\hbar\partial_t \bar{z}/l$ to be much smaller than the valley splitting $\sim 0.1$~meV. 
We emphasize that because $\hbar\omega_o\gg \hbar|\pmb{\omega}_B(z)|\sim 0.01$~meV, in our adiabatically moving quantum dots, the dynamics of spin does not need to be adiabatic with respect to the Zeeman field and we anticipate that resonance processes  with $\partial_t \bar{z}/l\sim|\pmb{\omega}_B(z)|$ can further enhance the fidelity of spin shuttling, see Sec.~\ref{sec:Dyn}.

\section{\label{sec:Zeeman} Inhomogeneous Zeeman field}
\subsection{Deterministic spin dynamics}
We first focus on a spin moving in an inhomogeneous Zeeman field and neglect for the moment the effect of SOI, i.e. $\pmb{\textbf{v}}=0$ in the Hamiltonian $H_\text{1D}$ of Eq.~\eqref{eq:H_1D_general}.
This simple case captures the response of planar hole nanostructures and of electron spins moving in micromagnetic fields and shows how the spin dynamics during shuttling can filter out the relevant low-frequency noise sources.

Assuming that the confinement potential is strong compared to the local Zeeman field and restricting for now to shuttling processes that are adiabatic compared to both orbital and spin dynamics, i.e. $\omega_o\gg | \pmb{\omega}_B(z)|\gg \partial_t \bar{z}/l$, we find by conventional time-dependent perturbation theory that the spin degree of freedom evolves according to the inhomogeneous Zeeman Hamiltonian
\begin{subequations}
\label{eq:H_IZ}
\begin{align}
 H_Z&=\frac{\hbar }{2} \pmb{\omega}_B[\bar{z}(t)] \cdot \pmb{\sigma} \ , \\ \pmb{\omega}_B[\bar{z}]&=\int dz |\psi(z-\bar{z})|^2 \tilde{\pmb{\omega}}_B(z) \ ,
\end{align}
\end{subequations}
see Appendix~\ref{App:Non_adiabatic} for more details.
The Zeeman energy of the quantum dot $\pmb{\omega}_B$ contains quantitative corrections coming from the inhomogeneity of the field averaged over the charge density   $|\psi(z-\bar{z})|^2\approx e^{-(z-\bar{z})^2/l^2}/l\sqrt{\pi}$ of the particle.
The adiabatic condition on the spin degrees of freedom constrains the shuttling velocity to be $\partial_t \bar{z}\ll \text{min}|\pmb{\omega}_B| l $. For typical values of $|\pmb{\omega}_B|/2\pi \sim 1- 10$~GHz and $l\sim 10-100$~nm, this condition is well satisfied for reasonable velocities $\lesssim 10$~m/s. We will further relax this condition in Sec.~\ref{sec:Dyn}.

The time-evolution of the spin generated by $H_Z$ is well approximated  by the unitary operator
\begin{equation}
\label{eq:U_B}
    U_Z(t)\approx e^{-i \theta_B[\bar{z}(t)] \textbf{n}_B
   [\bar{z}(t)]\cdot\pmb{\sigma}/2}e^{-i \Phi_B(t) \sigma_3/2} \ .
\end{equation}
The first transformation $e^{-i \theta_B \textbf{n}_B\cdot\pmb{\sigma}/2}$ locally diagonalizes $H_Z$ at position $\bar{z}$. 
The local angle $\theta_B[\bar{z}]$ and unit vector $\textbf{n}_B[\bar{z}]$ are  found explicitly solving the equation $\pmb{\omega}_B/|\pmb{\omega}_B|= \hat{R}_{B}(\theta_B)\textbf{n}_3$, for each value of $\bar{z}$. Here, $\textbf{n}_3=(0,0,1)$ and $\hat{R}_{B}(\theta_B)$ is  an anticlockwise rotation matrix around the axis $\textbf{n}_B$ of an angle $\theta_B$, see Appendix~\ref{sec:Rot} for more details and for a general solution for the vector $\textbf{n}_B$ and angle $\theta_B$. We conventionally choose the local angle $\theta_B$ to satisfy $\theta_B[\bar{z}=0]=0$ and $U_Z(t=0)=1$. 
Because of the adiabatic condition in the spin degrees of freedom, we discard negligible terms $\propto \partial_t \bar{z}/l$ generated by the first transformation, and the time-evolution in this locally rotated frame is the spin-dependent phase accumulation  given by the second exponential: $e^{-i \Phi_B(t) \sigma_3/2}$, with  $\Phi_B(t)=\int_0^t \left|\pmb{\omega}_B[\bar{z}(\tau)] \right|d\tau$. 
Non-adiabatic corrections to this model can prove beneficial for shuttling and are leveraged in Sec.~\ref{sec:Dyn}.

\subsection{\label{sec:shuttling-noise} Shuttling fidelity in a noisy environment}
The unitary operator $U_Z(t)$ in Eq.~\eqref{eq:U_B} describes the coherent deterministic time evolution of the spin. 
Because $U_Z$ can be characterized in experiments and can be compensated for or engineered to implement single-qubit gates, it does not influence the overall shuttling fidelity.
However, during shuttling the spin also experiences random fluctuations in the environment that result in a loss of its coherence. At small shuttling velocities, the dominant contribution in a conveyer-mode shuttling process is estimated to be the variation of spin splitting caused by charge noise~\cite{PRXQuantum.4.020305}. 
To describe this effect, we consider the noise Hamiltonian~\cite{PhysRevB.77.174509, 
PRXQuantum.2.010348,Piot2022, PhysRevLett.127.190501,PhysRevB.78.155329,PhysRevLett.105.266603}
\begin{equation}
\label{eq:H_N}
    H_N=\frac{\pmb{h}(t)\cdot\pmb{\sigma}}{2} \ ,
\end{equation}
where a stochastic, time-dependent vector $\pmb{h}(t)$ couples to the spin. Physically, this vector  originates from long-range fluctuations of the gate electric field (global noise sources) or from short-range atomistic defects (local noise sources) coupling to spin by the effective SOI or hyperfine interactions. This Hamiltonian can also describe the effect of small random variations of the trajectory of the shuttled spin in the inhomogeneous field. A detailed comparison between local and global noise sources is delayed to Sec.~\ref{sub: Examples}. We anticipate that, while the microscopic origin of the noise influences quantitatively the shuttling fidelity, the coherent spin dynamics  reduce the effect of the noise independently of the source, and for this reason we focus first on the simpler case of global noise sources. The derivations for general cases are provided in Appendix~\ref{app:SOI-homo-inhomo}.

In the interaction picture, $H_N$ is dressed by the time evolution of the spin as
\begin{subequations}
\begin{align}
  H_N^I&=  U_Z^\dagger H_NU_Z=\frac{1}{2}\pmb{h}(t)\cdot \hat{R}_Z(t) \pmb{\sigma} \ , \\
  \hat{R}_Z(t)&= \hat{R}_{B (t)}[\theta_B(t)]\hat{R}_{3}[\Phi_B(t)] \ ,
\end{align}
\end{subequations}
Here, $\hat{R}_Z$ is the combined rotation matrix generated by the transformation $U_Z$ and the notation $\hat{R}_{B(t)}[\theta_B(t)]$ emphasizes that $\hat{R}_{B}$ depends on time via its time-dependent rotation axis $\textbf{n}_B(t)$ and angle $\theta_B(t)$; $\hat{R}_3 $ is the rotation matrix about the local Zeeman axis, see Appendix~\ref{sec:Rot} for the explicit form.
When noise is small, $H_N$ generates the time-evolution operator $ U_N\approx e^{-i\pmb{\phi}_N(t)\cdot\pmb{\sigma}/2} $, with the vector of random phases
\begin{equation}
 \pmb{\phi}_N(t)=\frac{1}{\hbar}\int_0^t d\tau\pmb{h}(\tau)\hat{R}_{Z}(\tau)  \ .
\end{equation}

To quantify the error caused by the stochastic phase accumulation during shuttling, we introduce the fidelity of a single shuttling event
\begin{equation}
    \mathcal{F}=\frac{1}{2}\text{Tr}\left(U_\text{id}^\dagger U_{\text{re}}\right)=\frac{1}{2}\text{Tr}\left(e^{-i \pmb{\phi}_N \cdot\pmb{ \sigma}/2}\right)  \ ,
\end{equation}
that measures the distance between the ideal (coherent) and real (noisy) operations $U_\text{id}=U_Z$ and $U_{\text{re}}=U_ZU_N$, respectively.
The average shuttling fidelity $\bar{\mathcal{F}}$  is obtained by averaging  $\mathcal{F}$ over the probability distribution of $\pmb{\phi}_N$. 
Assuming a Gaussian-distributed noise~\cite{PhysRevB.77.174509,PRXQuantum.2.010348, PhysRevLett.127.190501,PhysRevB.78.155329,PhysRevLett.105.266603}, we obtain
\begin{equation}
\label{eq:averagine_F}
   \bar{\mathcal{F}}=\int_{-\infty}^{\infty}d\pmb{\phi}_N \frac{e^{-\pmb{\phi}_N\cdot\hat{\Sigma}^{-1}\pmb{\phi}_N/2}}{\sqrt{8\pi^3|\Sigma|}}\cos\left(\frac{\sqrt{\pmb{\phi}_N\cdot\pmb{\phi}_N}}{2 }\right) \ ,
\end{equation}
where we introduced the covariance matrix  
\begin{equation}
\label{eq:covariance}
    \hat{\Sigma}=\frac{1}{2\pi \hbar^2}\int_{-\infty}^{\infty} d\omega S(\omega) \hat{F}(\omega, t) \ ,
\end{equation}
with determinant $|\Sigma|$; $S(\omega)=\int dt e^{i\omega t} \langle h(t)h(0)\rangle $ is the power spectral function of the noise, which for simplicity we assumed to be isotropic and uncorrelated in space and spin directions, i.e., $\langle h_i(t)h_j(0)\rangle= \delta_{ij}\langle h(t)h(0)\rangle$. The generalization of Eq.~\eqref{eq:covariance} for noise sources that couple to the moving spin anisotropically~\cite{PhysRevLett.127.190501,PhysRevB.78.155329,PhysRevLett.105.266603} is straightforward and is provided in Appendix~\ref{app:SOI-homo-inhomo}.
The matrix of filter functions~\cite{PhysRevB.77.174509}
\begin{equation}
\label{eq:Filter-funct}
\hat{F}(\omega, t)=\int_{0}^{t} d\tau\int_{0}^{t} d\tau' e^{-i\omega(\tau-\tau')}\hat{R}_{Z}^T(\tau)\hat{R}_{Z}(\tau')
\end{equation}
depends on fast rotations around the local spin quantization axis [$R_{3}(\Phi_B)$], which account for the phase accumulated because of the Zeeman energy $|\pmb{\omega}_B|/2\pi\sim 10$~GHz,  and on slower rotations  $\sim 10-100$~MHz of the spin quantization axis [$R_B(\theta_B)$] caused by the motion of the spin in an inhomogeneous Zeeman field.

In realistic semiconducting devices, the spectral function $S(\omega)$ is strongly peaked at low frequencies and has a $1/\omega$ tail at large frequencies~\cite{camenzind2021spin, Yoneda2018,Struck2020}.
Because the transversal elements of $\hat{F}$ contain rapidly oscillating terms determined by $\Phi_B$, they are peaked at large frequencies in the GHz range, where the noise has less weight. For this reason, the dominant contribution to the fidelity arises from the longitudinal element of the covariance matrix $\hat{\Sigma}_{33}$, which is peaked at low frequencies, and is determined by the element $\hat{F}_{33}\equiv F$,
\begin{equation}
\label{eq:F33}
 {F}(\omega,t)=\int_{0}^{t} d\tau\int_{0}^{t} d\tau' e^{-i\omega(\tau-\tau')}\frac{\pmb{\omega}_B[\bar{z}(\tau)]}{|\pmb{\omega}_B[\bar{z}(\tau)]|} \cdot \frac{\pmb{\omega}_B[\bar{z}(\tau')]}{|\pmb{\omega}_B[\bar{z}(\tau')]|} \ ,
\end{equation}
of the matrix of filter functions $\hat{F}$~\footnote{ To derive Eq.~\eqref{eq:F33}, we used $[R_Z^T(\tau)R_Z(\tau')]_{33}=[R_B^T(\tau)R_B(\tau')]_{33}=(R_B(\tau) \textbf{n}_{3})\cdot(R_B(\tau') \textbf{n}_{3})=\pmb{\omega}_B[\bar{z}(\tau)]\cdot\pmb{\omega}_B[\bar{z}(\tau')]/|\pmb{\omega}_B[\bar{z}(\tau)]||\pmb{\omega}_B[\bar{z}(\tau')]|$.}.
In this case, the average shuttling fidelity becomes
\begin{equation}
\label{eq:average-shuttling}
\bar{\mathcal{F}} =e^{-\hat{\Sigma}_{33}/8}  \ .
\end{equation}
We note that the corrections coming from the fast-rotating transversal terms causing spin relaxation lead to a power-law decay, with slower time constants, instead of the faster exponential decay included here~\cite{bosco2023phase,PhysRevLett.127.190501,PhysRevB.78.155329,PhysRevLett.105.266603}.

\begin{figure}[t]
\centering
\includegraphics[width=0.49\textwidth]{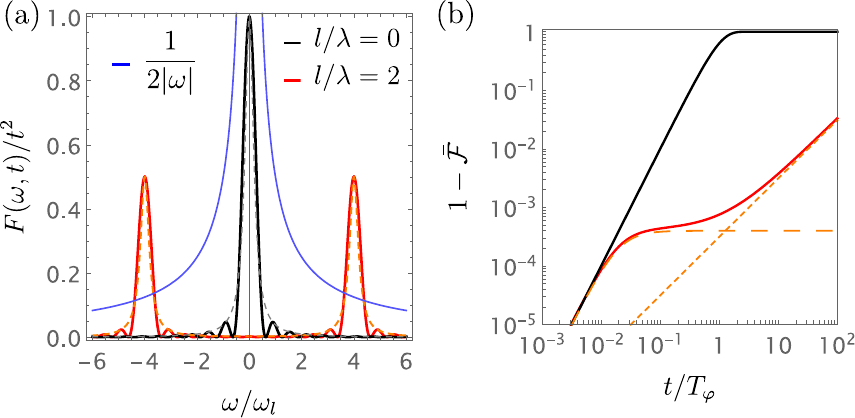}
\caption{Enhancing the shuttling fidelity by an inhomogeneous Zeeman field. We consider here global noise sources and the  Zeeman field in Eq.~\eqref{eq:IZ-rot}. (a) Filter functions. With a black and red curve, we show the filter functions of a shuttling experiment in a homogeneous  and an inhomogeneous Zeeman field rotating with period $\lambda=l/2$ [see Eq.~\eqref{eq:IZ-rot}], respectively. The solid lines show the exact filter functions given in Eqs.~\eqref{eq:f_FID} and ~\eqref{eq:F_FR_exact}, while the dashed lines show the approximated result in Eq.~\eqref{eq:F_FR}. For reference, we show with blue line a typical spectral function of the noise $S(\omega)\propto |\omega|^{-1}$.  We consider $\omega$ in units of $\omega_l=\bar{v}/l$ and we used $t=10/\omega_l$.
(b) Shuttling fidelity. Comparison of the infidelity $1-\bar{\mathcal{F}}$ in a doubly logarithmic plot. The time is measured in units of the dephasing time $T_\varphi$. A finite $\lambda$ improves the shuttling fidelity for global noise sources, as one can observe by comparing the black ($\lambda\to \infty$) and red ($\lambda=l/2$) curves. Solid lines show the exact results in Eqs.~\eqref{eq:fid_FID} and~\eqref{eq:Lorentzian_FR_decay} and dashed lines show the limiting cases in Eq.~\eqref{eq:Fid_FR}. We consider a noise with $\eta=0.01$ and $\omega_\lambda= \bar{v}/\lambda= 50/T_\varphi$, corresponding to $\omega_\lambda/2\pi\approx 8$~MHz for typical values $T_\varphi\approx 1$~$\mu$s.
\label{fig:filter_functions}
}
\end{figure}

Eqs.~\eqref{eq:F33} and~\eqref{eq:average-shuttling} highlight the fundamental role that the inhomogeneity of the Zeeman field has in determining the average shuttling fidelity $\bar{\mathcal{F}}$. In particular, the inhomogeneous tilt of the spin quantization axis encoded in the product $\pmb{\omega}_B[\bar{z}(\tau)]\cdot \pmb{\omega}_B[\bar{z}(\tau')]$ can substantially impact the filter function. We discuss this phenomenon in the next section by analysing a few key  examples. A comparison between the filter functions and average shuttling fidelities obtained for different cases is shown in Fig.~\ref{fig:filter_functions}.

\subsection{\label{sub: Examples} Suppressing noise by shuttling}
\subsubsection{\label{sub: Examples_homo} Spin rotation in homogeneous Zeeman fields} 
We consider first the simplest case where during shuttling the spin moves in a homogeneous Zeeman field, i.e. $\tilde{\pmb{\omega}}_B(z)=\pmb{\omega}_B(z)=\omega_B \textbf{n}_3$. This case is the aim of current experimental settings, but we will show that it does not always correspond to the highest shuttling fidelity.

If the Zeeman field does not depend on space the unitary time-evolution operator of the spin given in Eq.~\eqref{eq:U_B} reduces to the simple phase accumulation $U_Z=e^{-i \omega_B t\sigma_3/2}$, which rotates the spin around the fixed axis $\textbf{n}_3$.
Moreover, the product $\pmb{\omega}_B[\bar{z}(\tau)]\cdot \pmb{\omega}_B[\bar{z}(\tau')]= 1$ and the longitudinal filter function $F$ in Eq.~\eqref{eq:F33} simplifies to
\begin{equation}
\label{eq:f_FID}
    F(\omega,t)=\frac{4\sin^2(\omega t/2)}{\omega^2}\equiv F_\text{FID}(\omega,t) \ ,
\end{equation}
which corresponds to the filter function of a free-induction decay (FID) experiment~\cite{PhysRevB.77.174509}.

We remark that $F_\text{FID}$ is peaked at zero frequency $\omega=0$, where it grows as $F_\text{FID}(\omega=0,t)=t^2$, see the black line in Fig.~\ref{fig:filter_functions}(a). 
For this reason, the shuttling fidelity $\bar{\mathcal{F}}$, related to the longitudinal component $\hat{\Sigma}_{33}$ of the covariance matrix by Eq.~\eqref{eq:average-shuttling}, is determined by low-frequency noise which dominates the integral in Eq.~\eqref{eq:covariance}.

To explicitly compare different scenarios, we use here the typical spectral function measured in experiments~\cite{camenzind2021spin,Yoneda2018} 
\begin{equation}
\label{eq:S_1/f}
S(\omega)=\frac{2\pi\hbar^2}{T^{2-\eta}}\frac{1}{|\omega|^{1-\eta}} \ , 
\end{equation}
where $\eta\in (0,1]$ and we introduce the time scale $T>0$, that characterize the amplitude of the noise fluctuations in different experiments.
In particular, combining Eqs.~\eqref{eq:covariance},~\eqref{eq:average-shuttling}, and~\eqref{eq:f_FID}, we find that for FID, the average shuttling fidelity is
\begin{subequations}
\label{eq:fid_FID}
\begin{align}
  \bar{\mathcal{F}}_\text{FID}&=e^{-\left({t}/{T}\right)^{2-\eta} \cos \left({\pi  \eta }/{2}\right) \Gamma (\eta -2)/2}\approx e^{-{t^2}/{T_\varphi^2}} \ , \\
  \ T_\varphi&={2 T}\sqrt{\eta}  \ ,
\end{align}
\end{subequations}
where $\Gamma(x)$ is the gamma function.
The approximation reports the purely pink noise case, with $\eta\to 0^+$, such that the noise spectrum is $S(\omega)\propto 1/|\omega|$, see the blue line in Fig.~\ref{fig:filter_functions}(a).
Importantly, we stress that the dephasing time $T_\varphi\propto \sqrt{\eta}$ vanishes for purely  $1/|\omega|$ noise because of the characteristic non-integrable divergence at zero frequency.

The average shuttling infidelity $1-\bar{\mathcal{F}}$ for FID is shown with a black line in Fig.~\ref{fig:filter_functions}(b), and it will serve as a reference to compare different cases.
For typical experimental values of $T_\varphi \sim 1$~$\mu$s~\cite{Piot2022,hendrickx2020four,camenzind2021spin,Yoneda2018} and shuttling velocities of $\sim 1$~m/s, we obtain a loss of coherence of the spin within a distance $\bar{z}\sim 1$~$\mu$m.
Finally, we  remark here that as long as the motion of the spin remains adiabatic compared to orbital and Zeeman fields, the shuttling fidelity of FID is independent of the velocity of the quantum dot. This is not generally valid in the presence of inhomogeneity of the Zeeman field, as we discuss next.

\subsubsection{\label{sec:spin-prece}Spin precession in inhomogeneous Zeeman fields}
In striking contrast to the FID case, if the Zeeman field is inhomogeneous, the time-dependence of the product  $\pmb{\omega}_B[\bar{z}(\tau)]\cdot \pmb{\omega}_B[\bar{z}(\tau')]$ in Eq.~\eqref{eq:F33} shifts the weight of the longitudinal filter function $F$ to frequencies of tens of MHz, thus significantly improving the average shuttling fidelity. 

To illustrate this effect, we consider first a simple scenario where the moving spin precesses in the inhomogeneous Zeeman field 
\begin{equation}
\label{eq:IZ-rot_1}
 \frac{\tilde{\pmb{\omega}}_B^\text{P}(z)}{\tilde{\omega}_B} = \cos\left(\frac{2z}{\lambda}\right) \textbf{n}_3+\sin\left(\frac{2z}{\lambda}\right) \textbf{n}_2 =  \hat{R}_1\left(\frac{2z}{\lambda}\right) \textbf{n}_3 \ ,
\end{equation}
that fully rotates around a fixed axis.
The matrix $\hat{R}_1$ is reported in Appendix~\ref{sec:Rot} and describes a rotation around $\textbf{n}_1=(1,0,0)$ with period $\pi\lambda$. 

While being an ideal field, we emphasize that $\pmb{\omega}_B^\text{P}$ nicely describes a wide variety of devices. 
For example, in electronic systems $\pmb{\omega}_B^\text{P}$ matches the stray magnetic field produced by modular nanomagnets spaced by a distance $\pi \lambda$~\cite{10.1063/5.0139670}. 
In this case, we note that a small homogeneous magnetic field is required to polarize the magnets, but this field could be switched off after the initial polarization. 
Moreover, in planar hole nanostructures, $\tilde{\pmb{\omega}}_B^\text{P}$ reasonably approximates the strain- and electric field-induced tilting of the $g$ tensor~\cite{PhysRevLett.131.097002,PhysRevB.106.235426} caused by the periodic arrangement of gates required for a conveyer-mode shuttling architecture. For example, in neighbouring quantum dots defined in planar germanium heterostructures, $g$-tensors tilting of more than $40\%$~\cite{van2023coherent}, and even $g$-factors with opposite signs~\cite{PhysRevLett.128.126803}, have been recorded, suggesting that fully rotating fields as $\tilde{\pmb{\omega}}_B^\text{P}$ are within reach in these systems. 
A detailed discussion of the effects of residual homogeneous Zeeman fields is delayed to the next section~\ref{sec:SN}. 
We also anticipate that the field $\tilde{\pmb{\omega}}_B^\text{P}$ matches the effective Zeeman field produced by a finite SOI, as typical in hole nanowires and fin field effect transistors, as we will show in Sec.~\ref{sec:SOI}.

The Zeeman energy of the moving quantum dot appearing in $H_Z$ in Eq.~\eqref{eq:H_IZ} is
\begin{equation}
\label{eq:IZ-rot}
 \pmb{\omega}_B^\text{P}(\bar{z})= \omega_B  \hat{R}_1\left(\frac{2\bar{z}}{\lambda}\right) \textbf{n}_3  \ , \ \text{with} \ \ \omega_B= e^{-l^2/\lambda^2}\tilde{\omega}_B \ ,
\end{equation}
and is related to the local Zeeman energy $\tilde{\pmb{\omega}}^\text{P}_B(z)$ in $H_\text{1D}$ in Eq.~\eqref{eq:H_1D_general} by the well-known Gaussian renormalization factor $e^{-l^2/\lambda^2}$, which accounts for the effects of strong confinement and large inhomogeneity in the $z$-direction~\cite{PhysRevResearch.3.013081,PhysRevB.104.115425,PhysRevB.105.075308,PhysRevB.106.235408}.

In this case, the time-evolution operator in Eq.~\eqref{eq:U_B} and the kernel of the longitudinal filter function $F$ in Eq.~\eqref{eq:F33} reduce respectively to
\begin{subequations}
\label{UB_FR}
\begin{align}
     & U_Z^\text{P}(t)=e^{-i \bar{z}(t) \sigma_1/ \lambda}e^{-i \omega_B t \sigma_3/2} \ , \\  
      & \frac{\pmb{\omega}_B[\bar{z}(\tau)]}{|\pmb{\omega}_B[\bar{z}(\tau)]|} \cdot \frac{\pmb{\omega}_B[\bar{z}(\tau')]}{|\pmb{\omega}_B[\bar{z}(\tau')]|}=\cos\left[2\frac{\bar{z}(\tau)-\bar{z}(\tau')}{\lambda}\right] \ .  
\end{align}
\end{subequations}
In contrast to FID, in an inhomogeneous Zeeman field the quantum dot motion plays a critical role because the energetically favoured spin quantization axis varies at different positions and times. This results in spin precession during shuttling.

Considering a constant shuttling velocity $\bar{z}(t)=\bar{v}t$, the integral in Eq.~\eqref{eq:F33} defining $F$ can be evaluated exactly, and the complete solution is provided in Appendix~\ref{sec:Filter_funct}, see Eq.~\eqref{eq:Lorentzian_FR_decay}.
We find that an accurate approximation for the exact result is provided by the simple equation
\begin{equation}
\label{eq:F_FR}
F_\text{P}(\omega, t)\approx \frac{t^2}{2}\left[f_L\!\left(\frac{\omega-2\omega_\lambda}{2/t}\right)+f_L\!\left(\frac{\omega+2\omega_\lambda}{2/t}\right)\right] \ ,
\end{equation}
where $f_L(x)= (1+x^2)^{-1}$ is a Lorentzian function normalized as $f_L(0)=1$.
We introduce here the relevant frequency shift $\omega_\lambda= \bar{v}/\lambda$, quantifying the rate of change in spin quantization axis; in a similar way, we also define the frequency $\omega_l= \bar{v}/l$.

In Fig.~\ref{fig:filter_functions}(a), we show a comparison between the exact (solid lines) and the approximate (dashed lines). Importantly, $F_\text{P}$ comprised two functions peaked at finite frequencies $\pm 2\omega_\lambda$ and with broadening $1/t$ becoming narrower at large times.
Assuming an adiabatic shuttling velocity $\bar{v}= 1$~m/s and a typical gate pitch of $\pi\lambda=50$~nm, we find $\omega_\lambda/2\pi=10$~MHz, substantially shifting the relevant components of noise toward MHz frequencies, where the noise has lower weight (blue line). This shift is equivalent to an intrinsic dynamical decoupling  of the largest low-frequency noise.

Considering pink noise with the spectral function $S(\omega)$ in Eq.~\eqref{eq:S_1/f}, and using  Eqs.~\eqref{eq:covariance},~\eqref{eq:average-shuttling}, and~\eqref{eq:F_FR}, we can estimate the shuttling fidelity. 
The complete equation is provided in Eq.~\eqref{eq:F_FR_exact} and is shown with a solid red line in Fig.~\ref{fig:filter_functions}(b). For pure $1/|\omega|$ noise ($\eta\to 0^+$) this function can be approximated by
\begin{equation}
\label{eq:Fid_FR}
    \bar{\mathcal{F}}_\text{P}\approx  \left\{
\begin{array}{ll}
      e^{-t^2/T_\varphi^2 } ,        & t\lesssim 1/\omega_\lambda \\
      e^{-\omega_\lambda^2 T_\varphi^2 }  ,       & 1/\omega_\lambda \lesssim t \lesssim T_\text{P}=8  \omega_\lambda T^2/\pi    \\
     e^{-{t}/{T_\text{P}}}   ,             &  t\gtrsim T_\text{P}
\end{array} 
\right. \ ,
\end{equation}
and nicely matches the limiting behaviour of  $\bar{\mathcal{F}}_\text{P}$, see dashed lines in Fig.~\ref{fig:filter_functions}(b). Here, $T_\varphi$ is the FID dephasing time given in Eq.~\eqref{eq:fid_FID}.
At small values of $\omega_\lambda t$, corresponding to a few spin rotations during shuttling,  $ \bar{\mathcal{F}}_\text{P}\approx  \bar{\mathcal{F}}_\text{FID}$. However, if the spin experiences many rotations during shuttling and $\omega_\lambda T_\varphi \gtrsim 1$, the fidelity first saturates to a finite value following the interpolation function $\bar{\mathcal{F}}_\text{P}\approx e^{-t^2 f_L(\omega_\lambda t)/T_\varphi^2 }$, and then decays exponentially with a longer time constant $T_\text{P}$ that is \textit{independent} of the small diverging cut-off $\eta\to 0^+$.

Considering the estimated value of $\omega_\lambda/2\pi\approx 10$~MHz and $T_\varphi=1$~$\mu$s, we find a significant improvement in the shuttling fidelity by the inhomogeneous magnetic field compared to the FID, as shown in Fig.~\ref{fig:filter_functions}(b), with infidelities that remain below $10^{-3}$ for a much wider range of shuttling times. Because of the intrinsic  dynamical decoupling of low-frequency noise, the inhomogeneous Zeeman field  boosts the possible shuttling times to times a few orders of magnitude larger than the dephasing time $T_\varphi$, corresponding to a coherent shuttling over distances larger than 100~$\mu$m.

We also note that while we assumed for simplicity a constant absolute value of the Zeeman frequency $\omega_B$, see Eq.~\eqref{eq:IZ-rot}, because the term dominating the fidelity is independent of $\omega_B$, our results remain approximately valid also when $\omega_B$ has a spatial dependence, e.g. an additional oscillatory component with period $\pi\lambda$, provided that the minimal Zeeman frequency $\omega_B^\text{min}$ remains large compared to $\omega_\lambda$. More details on the effects of inhomogeneous $\omega_B (\bar{z})$ are provided in Sec.~\ref{sec:Dyn}.

\subsubsection{\label{sec:SN}Spin nutation in inhomogeneous Zeeman fields}

\begin{figure}[t]
\centering
\includegraphics[width=0.49\textwidth]{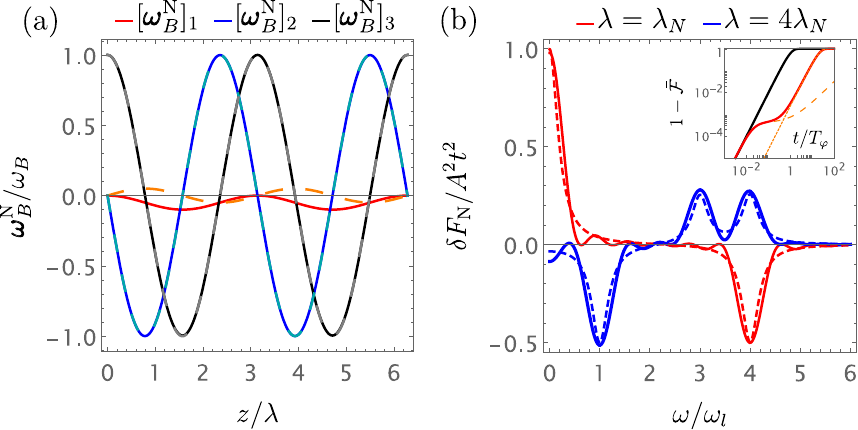}
\caption{Nutating Zeeman field. (a) Spatial dependence of the inhomogeneous  field. We show with red, blue, and black the three components of the field $\pmb{\omega}^\text{N}_B$. Solid [dashed] lines represent the out-of-phase [in-phase] nutation given by Eqs.~\eqref{eq:SNutation-Zeeman} and~\eqref{eq:inph-nut} [Eqs.~\eqref{eq:SNutation-Zeeman} and~\eqref{eq:oop-nut}]. The only component that is significantly modified in the two cases is the Zeeman field in the $x$-direction (red). We used here $A=0.05$ and $\lambda_N=\lambda$.
(b) Correction to the longitudinal filter function $\delta F$ caused by out-of-phase nutation. Red (blue) lines show the resonant (off-resonant) case  with $\lambda=\lambda_N= l/2$ ($\lambda=4\lambda_N= l/2$). Solid lines show the exact solution for $\delta F_\text{N}$, while the dashed lines are the approximation in Eq.~\eqref{eq:F_nutation}. We use here $t=10/\omega_l$. In the double logarithmic plot in the inset, we  show with a red solid line the infidelity of shuttling in an out-of-phase resonant nutating Zeeman field, see Eq.~\eqref{eq:fid_nutation}. The black and dashed orange lines show for reference the infidelity for a FID and a precessing field, see Fig.~\ref{fig:filter_functions}(b). The dotted orange line represents the contribution of the additional dephasing time $T_\varphi^\text{N}$. Here, $A=0.05$, $\lambda=\lambda_N=l/2$, and $\omega_\lambda=50/T_\varphi$.
\label{fig:filter_functions_Nutation}
}
\end{figure}

We now show that the enhancement of fidelity by inhomogeneous Zeeman field occurs in more general cases. In particular, we study here the nutating dynamics of a moving spin in the Zeeman field 
\begin{equation}
\label{eq:SNutation-Zeeman}
\pmb{\omega}_B^\text{N}(\bar{z}) =\omega_B  \hat{R}_{N}\!\left(\frac{2\bar{z}}{{\lambda}}\right)\textbf{n}_3 \ ,
\end{equation}
that rotates around an inhomogeneous vector.
The matrix $\hat{R}_{N}$ describes a general rotation around the oscillating unit vector
\begin{equation}
\label{eq:inph-nut}
\textbf{n}_\text{N}(z)=\frac{\textbf{n}_1}{\sqrt{1+A^2}}+ \frac{A \left[\cos\left({2\bar{z}}/{\lambda_N}\right) \textbf{n}_3 - \sin\left({2\bar{z}}/{\lambda_N}\right) \textbf{n}_2\right]}{\sqrt{1+A^2}}  \ .
\end{equation}
We refer to this process as to a nutation  out-of-phase because the rotation of $\textbf{n}_\text{N}$ is out-of-phase compared to the precessing Zeeman field, see Eq.~\eqref{eq:IZ-rot_1}. The amplitude of the nutation is characterizes by the dimensionless constant $A$ and by its period $\lambda_N$, which does not need to match the period $\lambda$ of the precession. We also only consider the cases where $\omega_\lambda \sim \omega_N \ll \omega_B$, with $\omega_N= \bar{v}/\lambda_N$.

Using Eq.~\eqref{eq:general_rot}, we can easily evaluate $\pmb{\omega}_B^\text{N}(\bar{z})$.
The components of the out-of-phase nutating Zeeman field are shown with solid lines in Fig.~\ref{fig:filter_functions_Nutation}(a). 
Compared to the rotating Zeeman field $\pmb{\omega}^\text{P}_B$ in Eq.~\eqref{eq:IZ-rot}, $\pmb{\omega}_B^\text{N}$ includes an additional component oscillating in the $x$-direction (red line). 
This oscillating term produces on average the finite homogeneous Zeeman field $-\omega_B A/(1+A^2)$, and thus $\pmb{\omega}_B^\text{N}$ nicely describes the effects of residual homogeneous fields in realistic experiments. These fields can occur because of non-zero polarizing magnetic field for electronic systems with nanomagnets~\cite{10.1063/5.0139670} or non-fully precessing $g$ tensors in hole nanostructures~\cite{van2023coherent}.
Here, we restrict ourselves to the case $A\ll 1$ and we show that in this case the shuttling fidelity is still strongly enhanced, however, we anticipate that similarly high fidelities can be engineered by increasing $\bar{v}$ also when residual homogeneous field is large, as we discuss in detail in Sec.~\ref{sec:Dyn}.

The spin dynamics in this case is well-approximated by the time-evolution operator
\begin{equation}
\label{eq:UB_FR_SN}
U_Z^\text{N}(t)=e^{-i \bar{z}(t) \textbf{n}_N[2\bar{z}(t)/\lambda_N]\cdot\pmb{\sigma} / \lambda}e^{-i \omega_B t \sigma_3/2} \ ,
\end{equation}
describing a spin nutation.
The longitudinal filter function $F$ in Eq.~\eqref{eq:F33} can be evaluated numerically. 
In the limit of small $A$, we find that  $ F_\text{N}=  F_\text{P}- \delta F_\text{N}$, with
\begin{multline}
\label{eq:F_nutation}
   \delta F_\text{N}\approx \frac{A^2t^2}{4}\left[2f_L\!\left(\frac{\omega-2\omega_\lambda}{2/t}\right) - f_L\!\left(\frac{\omega-2\omega_N}{2/t}\right) \right. \\
  \left. -f_L\!\left(\frac{\omega-2\omega_\lambda+2\omega_N}{2/t}\right) \right]  + (\omega\to -\omega) \ .
\end{multline}
Here, the notation $ \omega\to -\omega $ indicates that in the brackets there are three additional Lorentzian peaks obtained from the ones reported by inverting the frequency, and we neglected corrections $ \mathcal{O}(A^4) $ and combing from oscillations at higher frequencies.

The corrections $\delta F_\text{N}$  to the precessing filter function $F_\text{P}$ coming from out-of-phase nutation are shown with red and blue lines in Fig.~\ref{fig:filter_functions_Nutation}(b) for different values of $\lambda/\lambda_N$. We observe a good agreement of the approximated Eq.~\eqref{eq:F_nutation} (dashed lines) with the exact solution (solid lines).
Importantly, nutation introduces sideband peaks at frequencies $\omega=\pm 2 \omega_N$ and $\omega=\pm 2(\omega_\lambda-\omega_N)$ with amplitude $\propto A^2$.
When the period of nutation $\lambda_N$ is much shorter than the period of precession $\lambda$, and $\lambda_N\lesssim \lambda/2$ these sideband peaks sample noise at high frequency yielding negligible corrections to $F_\text{P}$ (blue lines).
In contrast, when $\lambda_N\gtrsim \lambda/2$, the sideband peaks of $\delta F_\text{N}$ occur at low frequencies. 
This effect results in a resonant condition at $\lambda_N=\lambda$, where the side peaks merge into the Lorentzian peak  $A^2t^2f_L(\omega t/2)$ sampling the noise at $\omega=0$ (red lines).

In this resonant scenario, and for the $1/|\omega|$ noise given in Eq.~\eqref{eq:S_1/f}, the average shuttling fidelity acquires a significant correction and becomes
\begin{equation}
\label{eq:fid_nutation}
    \mathcal{F}_\text{N}\approx   \mathcal{F}_\text{P}e^{-t^2/T_\text{N}^2}\ , \ \text{with} \ T_\text{N}=T_\varphi/A \ .
\end{equation}
This fidelity is shown in the inset of Fig.~\ref{fig:filter_functions_Nutation}(b).
Comparing to the dephasing time $T_\varphi$ in Eq.~\eqref{eq:fid_FID}, we observe that the time constant of the Gaussian decay is enhanced by the small amplitude of the nutation $A$. This decay time dominates the fidelity in the long time asymptotic.

The dependence of $T_\text{N}$ on $A$ can be understood in general by considering that at $\lambda=\lambda_N$ the out-of-phase nutating Zeeman field in Eq.~\eqref{eq:SNutation-Zeeman} contains on average the homogeneous component $-A \textbf{n}_1/(1+A^2)\approx -A^2 \textbf{n}_1$ along the main precession axis. This residual homogeneous field causes a constant dephasing during shuttling with time constant $T_\varphi (1+A^2)/ A = T_\text{N}+\mathcal{O}(A^2)$.
This interpretation clearly shows that when the spin degree of freedom is moved adiabatiacally compared to the Zeeman energy, the maximal enhancement of coherence occurs for effective inhomogeneous Zeeman fields that fully rotate during shuttling. 

We emphasize that the worst-case scenario presented here, where $\lambda=\lambda_N$, also requires the nutation in Eq.~\eqref{eq:SNutation-Zeeman} to be out-of-phase.
When the nutation is in-phase and is generated for example by the vector
\begin{equation}
\label{eq:oop-nut}
\textbf{n}_\text{N}(z)=\frac{\textbf{n}_1}{\sqrt{1+A^2}}+ \frac{A \left[\cos\left({2\bar{z}}/{\lambda_N}\right) \textbf{n}_3 + \sin\left({2\bar{z}}/{\lambda_N}\right) \textbf{n}_2\right]}{\sqrt{1+A^2}}  \ ,
\end{equation}
there is on average no homogeneous Zeeman field along the main precession axis, see the dashed lines in Fig.~\ref{fig:filter_functions_Nutation}(a), and thus $T_\text{N}\to \infty$.

\subsection{Local noise sources}
The noise model introduced in Eq.~\eqref{eq:H_N} assumes that during the shuttling the spin experience a random time-dependent Zeeman field $\pmb{h}(t)$ that is homogeneous in space. This model describes global noise sources originating for example from the fluctuation of the externally applied magnetic field or  long-range electric fields. 
Here, we analyse the effect of an inhomogeneous noise distribution during shuttling. We focus, in particular, on an ensemble of short-range impurities at fixed positions $z=z_k$, that couple to the spin via local interactions $\pmb{h}_k(t)$. This model describes well nuclear spins and local dynamical charge-traps electrostatically coupled to the dot.

In this case, the local noise Hamiltonian is
\begin{equation}
    \label{eq:H_N_inhomo}
    H_{N}= \frac{1}{2 n_0}\sum_k \delta(z-z_k)\pmb{h}_k(t)\cdot\pmb{\sigma} \ ,
\end{equation}
where $n_0$ is the atomic density and $\delta(z)$ is the delta function. 
The spin confined in the moving quantum dot has a charge density $|\psi[z-\bar{z}(t)]|^2\approx e^{-[z-\bar{z}(t)]^2/l^2}/l\sqrt{\pi}$ and experiences the time-dependent noise
\begin{equation}
    \label{eq:H_N_inhomo_spin}
    H^{L}_N= \frac{1}{2 n_0}\sum_k |\psi[z_k-\bar{z}(t)]|^2\pmb{h}_k(t)\cdot\pmb{\sigma} \ .
\end{equation}

Proceeding as in Sec.~\ref{sec:shuttling-noise}, assuming an isotropic and spatially uncorrelated noise with $\langle  \pmb{h}_k^n(t)\pmb{h}_{k'}^m(0)\rangle =\delta_{kk'}\delta_{nm}\int d\omega e^{-i\omega t} S(\omega)/2\pi$, and using the envelope function approximation $\sum_k\to \nu n_0\int dz $, where $\nu$ is the average percentage of defects,  we find that for local noise sources the longitudinal component of the filter function modifies as
\begin{multline}
\label{eq:F33I}
F^L(\omega, t)= \frac{\nu}{N}\int_{0}^{t} d\tau \int_{0}^{t}  d\tau'  e^{-i\omega(\tau-\tau')}  \\  \times e^{-\frac{[\bar{z}(\tau)-\bar{z}(\tau')]^2}{2l^2}}  \frac{\pmb{\omega}_B[\bar{z}(\tau)]}{|\pmb{\omega}_B[\bar{z}(\tau)]|} \cdot \frac{\pmb{\omega}_B[\bar{z}(\tau')]}{|\pmb{\omega}_B[\bar{z}(\tau')]|}
\end{multline}
where $N=\sqrt{2\pi}l n_0$ is the number of atoms in the dot. More detailed derivations of $F^L$, also including more general noise sources are provided in Appendix~\ref{app:SOI-homo-inhomo}.
Importantly, for local noise sources, the kernel of the filter function includes the additional weight $e^{-{[\bar{z}(\tau)-\bar{z}(\tau')]^2}/2{l^2}} $ that accounts for the locality of the noise and the spatial distribution of the spin. This term describes the motional narrowing of inhomogeneous noise during shuttling~\cite{PRXQuantum.4.020305}.

\begin{figure}[t]
\centering
\includegraphics[width=0.49\textwidth]{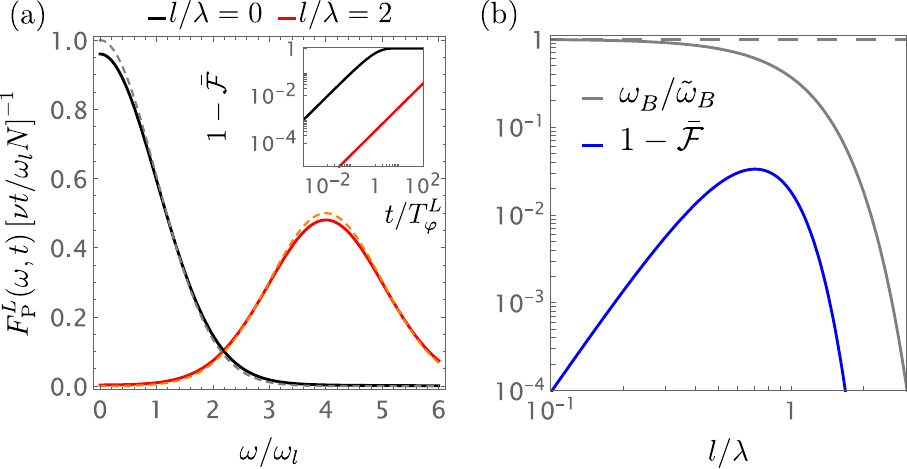}
\caption{Local noise sources. (a) 
Filter function and average shuttling fidelity for spin precessing in an inhomogeneous Zeeman field. 
Red and black curves represent the filter functions $F_\text{P}^L$ obtained at $l/\lambda=0$ and $l/\lambda=2$, respectively. Solid [dashed] lines show the exact [approximated] solution in Eq.~\eqref{eq:Lorentzian_FR_decay_in} [Eq.~\eqref{eq:F_FR,I}]. We used here $t=20/\omega_l$. In the inset, we show with a double logarithmic plot their corresponding average shuttling infidelity as a function of time, see Eq.~\eqref{eq:inf-INhomo_g}.
(b) Dependence of the shuttling infidelity and of the Zeeman energy on the inhomogeneity of the Zeeman field. In this double logarithmic plot, the blue curve represents the average shuttling as a function of $l/\lambda$ at the fixed time $t=T_0$ obtained by combining Eqs.~\eqref{eq:inf-INhomo_g} and~\eqref{eq:charge-noise-eq-tl}. With gray lines, we show the corresponding Zeeman energy $\omega_B^\text{P}$. The solid line represents the Zeeman energy $\pmb{\omega}_B^\text{P}$ in Eq.~\eqref{eq:IZ-rot}, which is renormalized by $e^{-l^2/\lambda^2}$ and obtained for a dot moving adiabatically compared to the Zeeman field, while the dashed line represents the Zeeman energy $\pmb{\omega}_B^\text{D}$ in Eq.~\eqref{eq:Zeeman-DD} when this adiabaticity condition is lifted. 
\label{fig:filter_functions_inhomogeneous}
}
\end{figure}

To illustrate its effect explicitly, we consider here the precessing Zeeman field $\pmb{\omega}_B^\text{P}$ given in Eq.~\eqref{eq:IZ-rot}. 
The coherent dynamics of the spin is not altered and the spin precesses according to the time-evolution operator $U_Z^\text{P}$ in Eq.\eqref{UB_FR}. However, the longitudinal filter function $F_\text{P}^L$ is significantly modified. By combing Eqs.~\eqref{eq:IZ-rot} and~\eqref{eq:F33I}, we derive an exact solution, reported in Eq.~\eqref{eq:Lorentzian_FR_decay_in}. In analogy to the global noise solution $F_\text{P}$ in Eq.~\eqref{eq:F_FR}, we find that $F_\text{P}^L$ can be approximated by
\begin{equation}
\label{eq:F_FR,I}
    F_\text{P}^L\approx \frac{\nu}{N}\frac{t}{2\omega_l }\left[f_G\!\left(\frac{\omega-2\omega_\lambda}{\omega_l}\right) +f_G\!\left(\frac{\omega+2\omega_\lambda}{\omega_l}\right) \right] \ ,
\end{equation}
with $f_G(x)=e^{-x^2/2}$ being a Gaussian normalized to $f_G(0)=1$.
As shown in Fig.~\ref{fig:filter_functions_inhomogeneous}(a), we observe a good match between the exact and the approximated solutions (solid and dashed lines, respectively). 

While qualitatively  $F_\text{P}^L$ and $F_\text{P}$ show a similar behaviour with the peaks of the filter function being shifted by the finite $\lambda$ to the higher frequencies $\pm 2\omega_\lambda$, with $\omega_\lambda=\bar{v}/\lambda$, we emphasize that there are a number of key differences between the two cases, see Eqs.~\eqref{eq:F_FR} and~\eqref{eq:F_FR,I}.
First, for local noise the peaks of $ F_\text{P}^L$ have a Gaussian lineshape that originates from the approximated charge density of the quantum dot $|\psi(z)|^2$ in contrast to the Lorentzian peaks of $ F_\text{P}$. Moreover, the broadening of the Gaussian peaks of $ F_\text{P}^L$ is time-independent and it is determined by the characteristic frequency $\omega_l=\bar{v}/l$. Finally, we observe that $ F_\text{P}^L\propto t/\omega_l$, while the for global noise the  $ F_\text{P}\propto t^2$, thus strongly impacting the average shuttling fidelity $ \bar{\mathcal{F}}$.

By considering the noise spectrum $S(\omega)$ in Eq.~\eqref{eq:S_1/f}, we find that  for local noise sources 
\begin{subequations}
\label{eq:inf-INhomo_g}
\begin{align}
  \bar{\mathcal{F}}_\text{P}^L&=e^{-  2^{\frac{\eta-6}{2}} \frac{\nu t}{N T} \Gamma \left(\frac{\eta }{2}\right) \left({\omega_l T}\right)^{\eta -1} \, _1F_1\left(\frac{1-\eta }{2};\frac{1}{2};\frac{-2l^2}{\lambda ^2}\right)}\approx e^{-{t}/{T_\varphi^L}} \ , \\  
  T_\varphi^L&=\omega_l \left(\sqrt{\frac{N}{\nu}} T_\varphi\right)^2  e^{\frac{2l^2}{\lambda^2}}  \ ,
\end{align}
\end{subequations}
where $\, _1F_1(a;b;c)$ is the hypergeometric function. We note that when the quantum dot is static, the FID dephasing times due to global and local noise are $T_\varphi=2T\sqrt{\nu}$ [see Eq.~\eqref{eq:fid_FID}] and $\sqrt{{N}/{\nu}}T_\varphi$, respectively, where the factor $\sqrt{{N}/{\nu}}$ accounts for the average percentage of local defects in the quantum dot~\cite{PhysRevLett.127.190501,PhysRevB.78.155329,PhysRevLett.105.266603}. 

In the inset of Fig.~\ref{fig:filter_functions_inhomogeneous}(a), we show the average shuttling infidelity for local noise sources, comparing the homogeneous $\lambda\to \infty$  (black curve) and precessing $\lambda=l/2$ (red curve)  Zeeman field cases.
In contrast to $\bar{\mathcal{F}}_\text{P}$ in Eq.~\eqref{eq:Fid_FR}, the shuttling fidelity $\bar{\mathcal{F}}_\text{P}^L$ follows an exponential decay with the time constant $T_\varphi^L$ being significantly larger than for global noise case because of the motional narrowing of local fluctuators. This effect can be clearly observed by considering  homogeneous Zeeman fields, and observing that for typical values of $\omega_l\sim 10-100$~MHz and $\sqrt{N/\nu}T_\varphi\sim 0.1-10$~$\mu$s,   $T_\varphi^L \gtrsim 10 \sqrt{N/\nu}T_\varphi$ also at $\lambda\to \infty$.

The additional  spin dynamics in the inhomogeneous field produces an additional beneficial effect which is  encoded in the Gaussian correction $e^{{2l^2}/{\lambda^2}}$ in Eq.~\eqref{eq:inf-INhomo_g}.
As a consequence, low-frequency local noise is substantially filtered out by the inhomogeneous field in  long quantum dots with $l\gtrsim \lambda$, see Fig.~\ref{fig:filter_functions_inhomogeneous}(a). 
However, we note that in long quantum dots the effective Zeeman energy $\omega_B$ is also reduced by a weaker Gaussian correction $e^{-{l^2}/{\lambda^2}}$, see Eq.~\eqref{eq:IZ-rot}, thus limiting the maximal values of the useful $l/\lambda$ ratio. 
This trade-off is highlighted in Fig.~\ref{fig:filter_functions_inhomogeneous}(b) by comparing solid blue and gray lines, that represent $T_\varphi^L$ and $\omega_B$, respectively.
Considering for example a typical gate pitch of $\pi\lambda\approx 50$~nm and realistic values of quantum dots length $l\approx 20$~nm, we observe a significant reduction of the noise with $T_\varphi^L\approx 20 T_\varphi^L(l/\lambda=0)$, still preserving a large Zeeman gap $\omega_B\approx 0.2 \tilde{\omega}_B$ at realistic values of magnetic fields $\sim 1$~T.
We anticipate that this trade-off between fidelity and Zeeman energy can be lifted by higher shuttling velocities that are not adiabatic with respect to the Zeeman energy, see the dashed gray curve, as we will discuss in Sec.~\ref{sec:Dyn}.

\subsection{Charge noise in inhomogeneous Zeeman fields}
We showed that an inhomogeneous Zeeman field dynamically decouples the moving spin from the dominant low-frequency noise, and thus provides an effective way to filter out the noise caused for example by hyperfine interactions with nuclear spins.
However, more care is required to analyse its effect on charge noise, because Zeeman field inhomogeneities characterized by the ratio $l/\lambda$ also render the spin susceptible to the fluctuations of the electrostatic environment, thus directly coupling the spin to these charge noise sources.
For this reason, current shuttling experiments minimize the inhomogeneity of the field and operate at $l/\lambda\ll 1$. 
We show here that while this approach indeed provides a coherent shuttling, the inhomogeneity-induced intrinsic dynamical decoupling also enables large shuttling fidelities at $l/\lambda\gtrsim 1$.
In particular, the time scale $T$ characterizing the noise spectral function $S(\omega)$ in Eq.~\eqref{eq:S_1/f} also depends on $l/\lambda$, thus further influencing the time $T_\varphi$.

To quantify this effect, we focus on the precessing spins discussed in Sec.~\ref{sec:spin-prece} and we include explicitly the coupling of the spin to charge noise due to the Gaussian renormalization of the Zeeman energy $\omega_B= e^{-l^2/\lambda^2}\tilde{\omega}_B$ given in Eq.~\eqref{eq:IZ-rot}.
Focusing on a local noise source labelled by $k$, small random  variations $\delta V_k(t)$ of the electrostatic environment cause fluctuations of the length $l$ and couple directly to the Zeeman energy resulting in the noise field~\cite{PRXQuantum.2.010348}
\begin{equation}
\pmb{h}_k(t)\approx \hbar {\pmb{\omega}}_B^\text{P}(z_k) \left[\frac{\partial_V \tilde{\omega}_B}{\tilde{\omega}_B} -\frac{2l^2}{\lambda^2} \frac{\partial_V l}{l} \right]\delta V_k(t)  \ .
\end{equation}
Here, we introduced the susceptibilities $\partial_V l $ and $\partial_V \tilde{\omega}_B$ of the length $l$ and of the local Zeeman field $\tilde{\omega}_B$ to variations in the environment. 
Moreover, we assumed that charge defects have a local effect on the spin, see Eq.~\eqref{eq:H_N_inhomo_spin}, however we point out that a similar noise Hamiltonian can be derived for global noise sources; corrections coming from intermediate-range noise are discussed in Appendix~\ref{app:SOI-homo-inhomo}.

Introducing now the pure $1/f$ charge noise spectral density $S_{\delta V}(\omega)= \bar{V}^2/|\omega|$, such that $\langle\delta V_k(t)\delta V_{k'}(0)\rangle =\delta_{k k'}\int d\omega S_{\delta V}(\omega) /2\pi$, we find the functional dependence of the time scale $T$ in Eq.~\eqref{eq:S_1/f} to be
\begin{equation}
\label{eq:T-charge}
T= \frac{\sqrt{2\pi}}{\bar{V}\tilde{\omega}_Be^{-l^2/\lambda^2}} \left|\frac{\partial_V \tilde{\omega}_B}{\tilde{\omega}_B} -\frac{2l^2}{\lambda^2} \frac{\partial_V l}{l} \right|^{-1}  \ .
\end{equation}
Away from sweet spots where the $T\to \infty$~\cite{PRXQuantum.2.010348} and by combining Eqs.~\eqref{eq:inf-INhomo_g} and~\eqref{eq:T-charge} we find
\begin{equation}
\label{eq:charge-noise-eq-tl}
 T_\varphi^L\approx  T_0 \frac{\lambda^4}{l^4}e^{\frac{4l^2}{\lambda^2}} \ , \ \ \text{with} \ \ T_0=  2\pi \eta\frac{N}{\nu }  \frac{\omega_l}{\tilde{\omega}_B^2} \frac{l^2}{\bar{V}^2(\partial_V l)^2} \ .
\end{equation}
We discarded here the term $\propto {\partial_V \tilde{\omega}_B}$ that is independent of $l/\lambda$ and is therefore clearly filtered out by the inhomogeneous field.

The functional dependence of the average shuttling fidelity on the inhomogeneity of the Zeeman field is illustrated in Fig.~\ref{fig:filter_functions_inhomogeneous}(b) with a blue curve.
As expected, for small values of $l/\lambda$ when the Zeeman field is rather homogeneous, the time constant $T_\varphi^L$ determining the shuttling fidelity decreases as $\propto l^4/\lambda^4$, resulting in an lower shuttling fidelity.  This power law is related to the typical scaling of the FID dephasing time $T_\varphi\propto l^2/\lambda^2\propto \sqrt{T_\varphi^L}$~\cite{PRXQuantum.2.010348}; we  note that also relaxation processes scale as the square of the inhomogeneity~\cite{PhysRevB.88.075301}.
In this regime, the noise is dominated by the variations of $\propto {\partial_V \tilde{\omega}_B}$ or by nuclear spin noise that are independent of $\lambda$.
However, if the Zeeman field inhomogeneity is large and $l/\lambda\gtrsim 1$, the induced intrinsic dynamical decoupling of the spin becomes effective and rapidly increases the shuttling fidelity. 
This same trend occurs also as a function of SOI, as we show in the following section.

\section{\label{sec:SOI} Spin-orbit interaction}
The SOI  causes a spin rotation depending on the velocity of the particle. 
This effect is captured by the term $\textbf{v}(z) $ in $H_\text{1D}$~\eqref{eq:H_1D_general} and is strongly enhanced in hole nanostructures, where the SOI are large and cause full spin rotations in lengths of a few tens of nanometers~\cite{camenzind2021spin, Wang2022, PhysRevResearch.3.013081, Froning2021, geyer2022two}.
We show here that SOI generally produces an inhomogeneous fully-rotating Zeeman field matching the ones analysed in Sec.~\ref{sec:Zeeman}.

To highlight the role of SOI, we rewrite Eq.~\eqref{eq:H_1D_general} as
\begin{equation}
\label{eq:H_1d_SOI}
H_\text{1D}=\frac{\left[p-m \textbf{v}(z)\cdot\pmb{\sigma}\right]^2}{2m}+U(z)+ \frac{\hbar\tilde{\pmb{\omega}}_B(z)\cdot\pmb{\sigma}}{2}\ ,
\end{equation}
with $U(z)=m[\omega_o^2(z-\bar{z})^2-|\textbf{v}(z)|^2]/2$.
We remove the SOI by the exact unitary transformation  
\begin{equation}
\label{eq:path-ordergauge}
 S= \mathcal{P}\text{exp}\left({i\frac{m}{\hbar}\int_{0}^z {ds}\textbf{v}(s)\cdot\pmb{\sigma}}\right)  
\end{equation} 
satisfying $S^\dagger [p-m \textbf{v}(z)\cdot\pmb{\sigma}] S= p$, and where $\mathcal{P}\text{exp}$ is the path-ordered exponential. 
Generally, $S$ describes an inhomogeneous spin rotation around a local axis. 

To find an explicit expression for this rotation, we restrict our analysis to SOI of the form 
\begin{equation}
\textbf{v}(z)=v_s \textbf{n}_{s} +\delta\textbf{v}(z) \ .
\end{equation}
By introducing the SOI length $\lambda_{s}=\hbar/m v_s$, we find
\begin{equation}
\label{eq:trafo_S}
S=e^{i z\textbf{n}_s\cdot\pmb{\sigma}/\lambda_{s}} e^{i \pmb{\phi}_s(z)\cdot\pmb{\sigma}} \ .
\end{equation}
For sufficiently small $\delta \textbf{v}(z)/v_s$, i.e. the individual components of the inhomogeneous term are bounded by  $m\int_0^zds \delta\textbf{v}_j(s)/\hbar<\pi$, the phases $\pmb{\phi}_s(z)$ can be estimated by a second order Magnus expansion~\cite{ZEUCH2020168327} as
\begin{equation}
\label{eq:lambda_gen}
\pmb{\phi}_s(z)\approx\frac{m}{\hbar}\int_{0}^z ds \delta\tilde{\textbf{v}}(s)+ \frac{m^2}{\hbar^2}\int_{0}^{z} ds \int_{0}^{s} ds' \delta\tilde{\textbf{v}}(s)\times\delta\tilde{\textbf{v}}(s') \ ,
\end{equation}
with $\delta\tilde{\textbf{v}}(z)=\hat{R}_{s}(2z/\lambda_{s})\delta{\textbf{v}}(z)$; $\hat{R}_{s}$ is a rotation matrix by the fixed SOI axis $\textbf{n}_s$.
Here, the first integral term captures the effect of a varying amplitude of the SOI, while the second term captures the first correction due to a small tilting of the vector of SOI. 
We note that for SOI with a constant direction  $\textbf{v}(z)= v_s(z) \textbf{n}_s$, Eq.~\eqref{eq:lambda_gen} is exact and the second integral vanishes.

Projecting the transformed Hamiltonian onto the moving charge state of the quantum dot $|\psi(z-\bar{z})|^2$, we find a spin model $H_Z=\hbar\pmb{\omega}_B[\bar{z}(t)]\cdot\pmb{\sigma}/2$, analogous to Eq.~\eqref{eq:H_IZ}, with effective Zeeman field
\begin{equation}
\label{eq:omegaB_SOI}
    \pmb{\omega}_B[\bar{z}]=\int dz |\psi(z-\bar{z})|^2  \hat{R}_{ \delta(z)}[\phi_s(z)] \hat{R}_{s}[2z/\lambda_s] \tilde{\pmb{\omega}}_B(z) \ ,
\end{equation}
with $\pmb{\phi}_s(z)=\phi_s(z)\delta\textbf{n}(z)$; the matrix $\hat{R}_{ \delta(z)}$ describes a general rotation around the local axis $\delta\textbf{n}(z)$.
We now examine different cases. To highlight the effect of SOI, we restrict ourselves to the analysis of a homogeneous Zeeman field, and we consider an homogeneous Zeeman field $\tilde{\pmb{\omega}}_B(z)=\tilde{\pmb{\omega}}_B$.

\subsection{Spin precession in homogeneous SOI}
We first consider the homogeneous SOI 
\begin{equation}
\label{eq:vh}
\textbf{v}_\text{H}(z)= v_s\textbf{n}_\text{s}   \ .
\end{equation}
The effective Zeeman field then reduces to~\cite{PRXQuantum.2.010348}
\begin{equation}
\pmb{\omega}_B(z)=\tilde{\pmb{\omega}}_B^\parallel+ e^{-l^2/\lambda_s^2} \hat{R}_{s}(2z/\lambda_s) \tilde{\pmb{\omega}}_B^\perp\ ,
\end{equation}
where $\tilde{\pmb{\omega}}_B^{\perp,\parallel}$ are the component of the Zeeman field perpendicular and parallel to $\textbf{n}_s$, respectively.

If the SOI and Zeeman vectors are aligned [$\textbf{n}_s\parallel \tilde{\pmb{\omega}}_B$ and $ \pmb{\omega}_B(z)=\tilde{\pmb{\omega}}_B^\parallel$], the spin rotates around a fixed axis  and the noise filter function reduces to $F_\text{FID}$ in Eq.~\eqref{eq:f_FID} as discussed in Sec.~\ref{sub: Examples_homo}.
In contrast, if the  SOI and Zeeman vectors are perpendicular to each other [$\textbf{n}_s\perp \tilde{\pmb{\omega}}_B$ and $\pmb{\omega}_B(z)= e^{-l^2/\lambda_s^2} \hat{R}_{s}(2z/\lambda_s) \tilde{\pmb{\omega}}_B^\perp$], the spin precesses around an effective Zeeman field rotating around a fixed axis in analogy to $\pmb{\omega}_B^\text{P}$ in Eq.~\eqref{eq:IZ-rot}, see Sec.~\ref{sec:spin-prece}.
In this case, the period of the rotation of the effective Zeeman field is determined by the SOI length $\lambda_s$; the dynamics of the spin is then given by the time-evolution operator $U_Z^\text{P}$ in Eq.~\eqref{UB_FR}.

However, because of the transformation $S$ in Eq.~\eqref{eq:trafo_S}, the response of the system to noise differs from the one discussed in Sec.~\ref{sec:Zeeman}. 
In this case, there is an important difference between global and local noise sources. 
For the global noise modeled by $H_N$ in Eq.~\eqref{eq:H_N},  the transformation $S$ rotates the global stochastic vector $\pmb{h}$ as $\pmb{h}\to e^{-l^2/\lambda_s^2} \hat{R}_{s}(2\bar{z}/\lambda_s)  \pmb{h}$. Because this additional rotation compensates for the spin dynamics, the low-frequency noise is not filtered out and $F=e^{-2l^2/\lambda_s^2}F_\text{FID}$. This change results in the rescaling of the dephasing time $T_\varphi\to T_\varphi e^{l^2/\lambda_s^2}$, i.e., the dephasing time is increased inverse proportionally to the Zeeman energy renormalization, see Eq.~\eqref{eq:IZ-rot}.

In contrast, for local noise sources, $H_{N}^L$ in Eq.~\eqref{eq:H_N_inhomo_spin} transforms as $\pmb{h}_k\to  \hat{R}_{s}(2z_k/\lambda_s)\pmb{h}_k $. Because the rotations in this case are local, the noise response of this system is described by the longitudinal filter function $F^{L}_\text{P}$ in Eq.~\eqref{eq:F33I},  resulting in the average shuttling fidelity $\bar{\mathcal{F}}_\text{P}^L$  given in Eq.~\eqref{eq:inf-INhomo_g}.
More details on this different noise response, including a general derivation for intermediate-range noise are provided in Appendix~\ref{app:SOI-homo-inhomo}.

\subsection{Spin nutation in inhomogeneous SOI}

\begin{figure}[t]
\centering
\includegraphics[width=0.49\textwidth]{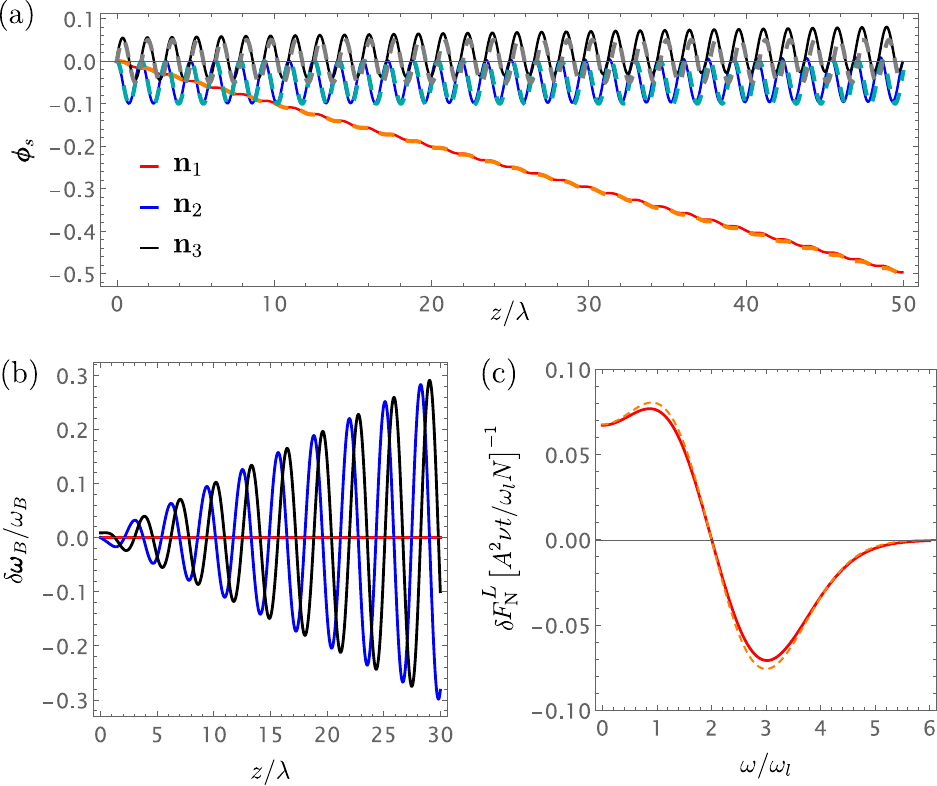}
\caption{Inhomogeneous SOI. (a) Inhomogeneous accumulated phases $\pmb{\phi}_s$ defining the transformation $S$ in Eq.~\eqref{eq:trafo_S}. We consider here the inhomogeneous SOI $\textbf{v}_\text{N}$ in Eq.~\eqref{eq:v_Nut} and we show with red, blue, and black curves the $\textbf{n}_1$, $\textbf{n}_2$, and $\textbf{n}_3$ component of $\pmb{\phi}_s$, respectively.  Solid [dashed] lines represent the exact [approximate] solution obtained by discretizing Eq.~\eqref{eq:path-ordergauge} [from Eq.~\eqref{eq:phi-s}]. We use  $A=0.2$ and $\lambda_s=\lambda_N=l$.
(b) Variation of the effective Zeeman field $\delta\pmb{{\omega}}_B$ caused by the inhomogeneous SOI. We show here the solution obtained combining Eqs.~\eqref{eq:omegaB_SOI} and~\eqref{eq:phi-s}, however we note that the simpler approximation provided in Eq.~\eqref{eq:change-zeeman} accurately reproduces the behaviour of $\delta \pmb{\omega}_B$. The color code and parameters used  are the same as in (a). 
(c) Correction of the filter function for local noise sources. We show with solid and dashed lines the exact and approximated [Eq.~\eqref{eq:FF-Soi}] solution of the longitudinal filter function evaluated at $t=20/\omega_l$.
\label{fig:filter_functions_SOI}
}
\end{figure}

Our general theory describes small variation of the SOI direction during shuttling. 
Such variations can arise in hole nanostructures for example in planar germanium and  silicon fin field-effect-transistors because of gate-induced strain and electric field modulations which can impact amplitude and direction of the SOI field~\cite{PhysRevB.104.115425,PhysRevB.105.075308,PhysRevB.106.235408,PhysRevB.106.235426,PhysRevLett.131.097002}.
These variations are captured by the additional phases $\pmb{\phi}_s$ in the transformation $S$, see Eq.~\eqref{eq:lambda_gen}.
To illustrate this effect, we consider a concrete example where the SOI precesses as 
\begin{equation}
\label{eq:v_Nut}
\textbf{v}_\text{N}(z)= v_s \left[\textbf{n}_1 - A \sin\left(\frac{2 z}{\lambda_N}\right)\textbf{n}_2+ A \cos\left(\frac{2 z}{\lambda_N}\right)\textbf{n}_3 \right] \ .
\end{equation}  
The precession of the SOI has a largely different effect than the precession of the inhomogeneous Zeeman field in Eq.~\eqref{eq:inph-nut}.

By using Eq.~\eqref{eq:lambda_gen}, we find that when $A\ll 1$ the inhomogneneous SOI leads to the phases
\begin{equation}
\label{eq:phi-s}
\pmb{\phi}_s\approx\frac{A^2 [\sin (2 k  z)-2 k  z]}{4k^2\lambda_s^2} \textbf{n}_1-\frac{A\sin ^2(k  z) }{k\lambda_s} \textbf{n}_2+\frac{A\sin (2 k  z) }{2 k \lambda_s} \textbf{n}_3 \ ,
\end{equation}
where we define the wavevector $k=1/\lambda_s+1/\lambda_N$. 
This equation remains rather accurate for large values of  $z \lesssim  1/kA^2$, as we show in Fig.~\ref{fig:filter_functions_SOI}(a) by comparing this approximation to the numerical integration of the path ordered exponential in Eq.~\eqref{eq:path-ordergauge}.
 
We focus on the homogeneous Zeeman field $\tilde{\pmb{\omega}}_B= \tilde{\omega}_B\textbf{n}_3$ that is perpendicular to the constant component of the SOI.
For simplicity, we now restrict ourselves to the case $\lambda_s=\lambda_N$; we will lift this fine-tuned condition later.
From Eq.~\eqref{eq:omegaB_SOI}, we find the effective Zeeman energy
 \begin{subequations}
 \label{eq:change-zeeman}
 \begin{align}
 \pmb{\omega}_B^\text{N} (\bar{z})&=  \omega_B \hat{R}_1\left(\frac{2\bar{z}}{\lambda}\right)+ \delta \pmb{\omega}_B (\bar{z}) \ , \\
 \frac{\delta\pmb{{\omega}}_B(\bar{z})}{\omega_B}&\approx 
 \frac{A^2 \bar{z} }{ 2k\lambda_s^2}\left[\cos(2\bar{z}/\lambda_s)\textbf{n}_2+\sin(2\bar{z}/\lambda_s)\textbf{n}_3 \right]\ .
 \end{align}
\end{subequations}
The first term in $\pmb{\omega}_B^\text{N}$ is equivalent to $\pmb{\omega}^\text{P}_B$ in Eq.~\eqref{eq:IZ-rot} and includes both the renormalization of Zeeman energy $\omega_B= e^{-l^2/\lambda_s^2}\tilde{\omega}_B$ and SOI-induced rotation $\hat{R}_s=\hat{R}_1$. 
The correction to the effective field $\delta\pmb{{\omega}}_B$ arising from the precession of SOI vector is shown in Fig.~\ref{fig:filter_functions_SOI}(b). We note that the largest correction originates from the term $\propto \bar{z} \textbf{n}_1$ of $\pmb{\phi}_s$ in Eq.~\eqref{eq:phi-s}, which increases linearly with $\bar{z}$, and produces the simple approximate expression provided in Eq.~\eqref{eq:change-zeeman}. 

Focusing on local noise sources, the additional local rotation of the Zeeman field $\delta\pmb{\omega}_B$ caused by the inhomogeneity of the SOI modifies the longitudinal filter function as
\begin{subequations}
\label{eq:FF-Soi}
\begin{align}
F_\text{N}^L&=F_\text{P}^L+ \delta F_\text{N}^L \ , \\
\delta F_\text{N}^L& =\frac{A^2 \nu t}{4\omega_l k \lambda_s^2 N}\left[f_G'\!\left(\frac{\omega-2\omega_\lambda}{\omega_l}\right) -f_G'\!\left(\frac{\omega+2\omega_\lambda}{\omega_l}\right) \right] \ , 
 \end{align}
\end{subequations}
where $F_\text{P}^L$ is given in Eq.~\eqref{eq:F_FR,I}, the frequency shift is $\omega_\lambda=\bar{v}/\lambda_s$, and we introduced the first derivative of the function $f_G$ as $f_G'(x)=-x e^{-x^2/2}$.

We show the variation $\delta F_\text{N}^L$ of the filter function caused by the inhomogeneous SOI in  Fig.~\ref{fig:filter_functions_SOI}(c).  Compared to the homogeneous SOI case, $F_\text{N}^L$ acquires only a small correction which scales with $A^2$ and is centered at $\omega\pm 2\omega_\lambda$. Interestingly, because of the linear increase of the Zeeman field $\propto \bar{z}$, the Gaussian shape of the peaks is modified by a polynomial correction. We anticipate that a similar polynomial renormalization appears also when the moving spin is resonantly driven, as we discuss in Sec.~\ref{sec:Dyn}.
We note that the corrections caused by the SOI precession are negligible in the regime considered. In contrast to the case of precessing Zeeman field and global noise discussed in Sec.~\ref{sec:SN}, they only quantitatively renormalise the exponential decay of shuttling fidelity.
In particular, the SOI precession renormalizes the decay rate as
\begin{equation}
\frac{1}{T_\varphi^L}\to\frac{1}{T_\varphi^L}\left(1+ \frac{A^2 l^2}{k \lambda_s^3}\right)= \frac{1}{T_\varphi^L}\left(1+ \frac{A^2 l^2}{2 \lambda_s^2}\right) \ ,
\end{equation}
where $T_\varphi^L$ is defined in Eq.~\eqref{eq:inf-INhomo_g}.

We now examine the case $\lambda_s\neq\lambda_N$, the inhomogeneous Zeeman field $\pmb{\omega}_B^\text{N}$ in Eq.~\eqref{eq:change-zeeman} acquires the additional correction
\begin{equation}
\frac{A \tilde{\omega}_B}{2 k \lambda_s}\left[e^{-l^2/\lambda_N^2} \cos (2 \bar{z}/\lambda_N)-e^{-l^2/\lambda_s^2} \cos (2 \bar{z}/\lambda_s)\right] \textbf{n}_1 \ ,
\end{equation}
that is linear in $A$ and is aligned to the homogeneous SOI direction. 
This term causes extra peaks in the longitudinal filter function 
\begin{equation}
\delta F_N^L=\frac{A^2 t\nu}{16\omega_l k \lambda_s^2 N }\left[f_G\!\left(\frac{\omega\pm 2\omega_N}{\omega_l}\right) -f_G\!\left(\frac{\omega\pm 2\omega_\lambda}{\omega_l}\right) \right] \ , 
\end{equation}
with $\omega_N=\bar{v}/\lambda_N$. 
These peaks are qualitatively similar to the ones in $F_\text{P}^L$ given in Eq.~\eqref{eq:F_FR,I}, and only provide an additional correction to the decay rate $\propto A^2$.

\section{\label{sec:Dyn} Resonant dynamical decoupling}
The average shuttling fidelity can be further enhanced by appropriately engineering the trajectory of the spin while shuttled. As anticipated, by rendering the quantum dot motion non-adiabatic with respect to Zeeman field, but still slow compared to the orbital splitting, the resonantly-induced deterministic spin dynamics more effectively filter out the low frequency noise, thus resulting in higher shuttling fidelities. 
In particular, we propose two different approaches: a fast time-modulation of the position of the quantum dot, and a fast shuttling in a weakly inhomogeneous Zeeman field. 

In this section, we restrict our analysis to shuttling experiments where the spin moves in a precessing inhomogeneous Zeeman fields and focus on local noise sources. As discussed in Sec.~\ref{sec:SOI}, this case is equivalent to a system with an homogeneous SOI. 

\subsection{\label{sec:FMP} Fast time-modulated position}

\begin{figure}[t]
\centering
\includegraphics[width=0.49\textwidth]{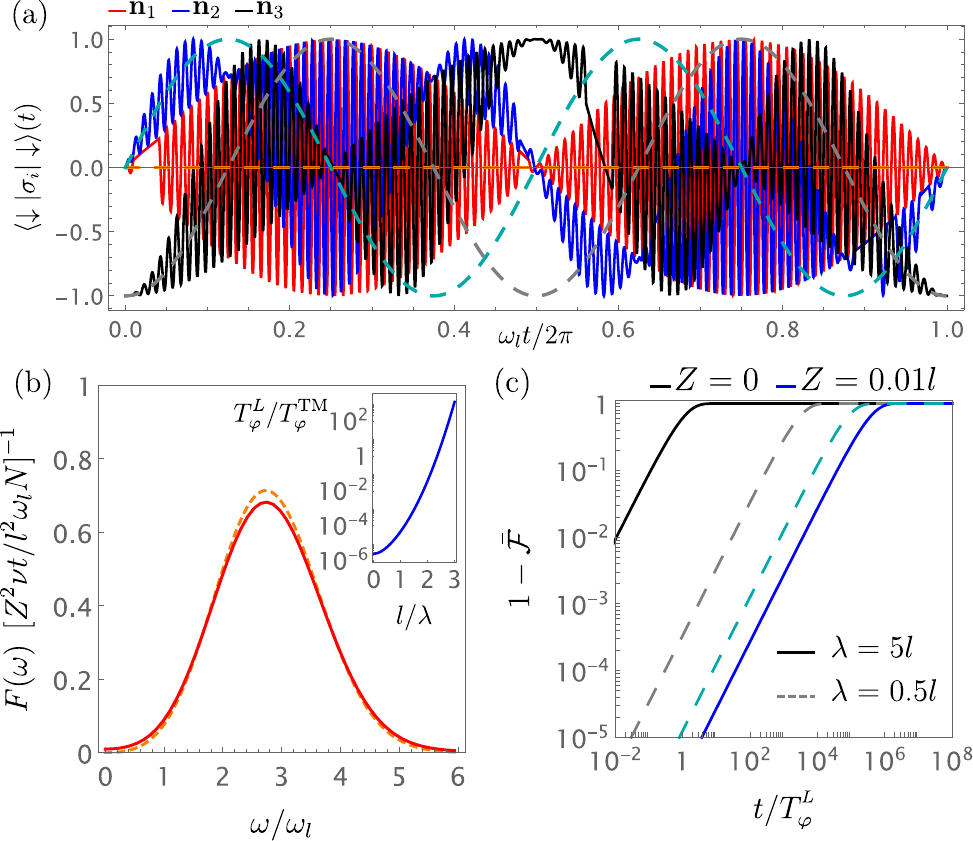}
\caption{Resonant dynamical decoupling with a time-modulated position. (a) Time evolution of the spin. We consider an initial spin state in the $|\downarrow\rangle$, which evolves according to the unitary time-evolution operator $U_\text{TM}$ in  Eq.~\eqref{eq:UTM}, and we show with red, blue, and black curves its expectation values of the spin operators $\pmb{\sigma}$ aligned to the $\textbf{n}_1$, $\textbf{n}_2$, and $\textbf{n}_3$ directions, respectively. Solid [dashed] lines show the cases $Z=0.01 l$ [$Z=0$]. We used $\lambda=l$ and $\omega_d= 100 \omega_l$. 
(b) Filter function. We show here the longitudinal filter function. The solid line represent the exact solution of the integral in Eq.~\eqref{eq:F33_D}, while the dashed line is the low-frequency approximated result in Eq.~\eqref{eq:F33_D_app}. In the inset, we show a logarithmic plot of the rate of decay $1/ T_\varphi^\text{TM}$ of the average shuttling fidelity against the inhomogeneity of the field $l/\lambda$, see Eq.~\eqref{eq:fid_TM}. We use the same values as in (a), $t=20/\omega_l$, and $\eta=0.1$.
(c) Average shuttling infidelity. With a double logarithmic plot we illustrate the enhanced fidelity obtained with the resonant time-modulation. Solid [dashed] lines are obtained for large [small] inhomogeneities with $\lambda=5 l$  [$\lambda=0.5 l$]. Blue and black lines show the fidelity with and without the additional resonant modulation, respectively. The parameters used are the same as above.
\label{fig:filter_functions_dynamical}
}
\end{figure}

\subsubsection{General solution}
We consider a time-modulated position of the quantum dot
\begin{equation}
\label{eq:TM-pos}
\bar{z}(t)=\bar{v} t+Z \cos(\omega_d t) \ ,
\end{equation}
which is modulated with an additional signal with amplitude $Z$ and frequency $\omega_d$.
We restrict to small resonant modulation with $Z\ll l$ and $\omega_d\sim \omega_B$.  
This additional driving term in the spin position can be experimentally achieved by appropriately designing the ac pulses of a conveyer-mode shuttler, and could be implemented in electronic systems with nanomagnets and in hole nanostructures. 

The additional small driving term induces resonant dynamics in the spin degrees of freedom, thus lifting the adiabaticity condition compared to the Zeeman energy discussed in Sec.~\ref{sec:Zeeman}, however because $\omega_B\ll \omega_o$, we still consider the motion to be adiabatic compared to the orbital degree of freedom. In particular, we note that this system is still well-described by the Hamiltonian $H_Z$ in Eq.~\eqref{eq:H_IZ}, up to small corrections of order $\omega_B^2/\omega_o$, that are derived in Appendix~\ref{App:Non_adiabatic}.
Because of the fast modulation, however, the adiabatic time evolution operator $U_Z$ provided in  Eq.~\eqref{eq:U_B} does not describe accurately the time-evolution of $H_Z$. 
In this case, the neglected dynamical term $i\hbar U_Z^\dagger \partial_t U_Z$ is relevant and induces additional resonant spin dynamics. 

By applying  the transformation $U_Z^\text{TM}=e^{-i\theta_B(\bar{z})\textbf{n}_B(\bar{z})\cdot\pmb{\sigma}/2}e^{-i\omega_d t\sigma_3/2}$ to $H_Z$ in Eq.~\eqref{eq:H_IZ}, we find the effective Hamiltonian
\begin{equation}
\label{eq:H_d}
H_\text{TM}=\frac{\hbar\Delta(\bar{{z}})}{2}\sigma_3- \frac{\hbar\partial_t \bar{z}}{2}\delta\pmb{\theta}_B(\bar{z}) \hat{R}_3(\omega_d t)\cdot\pmb{\sigma} \ .
\end{equation}
For convenience, here the second transformation in $U_Z^\text{TM}$ moves the system to a frame rotating at the frequency of the drive $\omega_d$ rather than at the Zeeman frequency as in $U_Z$ in Eq.~\eqref{eq:U_B}.
We also introduce the detuning $\Delta(\bar{z})=|\pmb{\omega}_B(\bar{z})|-\omega_d$ and the vector
\begin{multline}
\label{eq:theta_B^d}
\delta\pmb{\theta}_B=\theta_B'\textbf{n}_B +\theta_B (\textbf{n}_B\cdot\textbf{n}_B') \textbf{n}_B
 +\sin(\theta_B)(\textbf{n}_B\times\textbf{n}_B')\times\textbf{n}_B \\
+[1-\cos(\theta_B)] (\textbf{n}_B\times\textbf{n}_B') \ ,
\end{multline}
which is derived by using Eq.~\eqref{eq:Rot_general_vect} and the equality $e^{-X}(\partial_t e^{X})=\int_0^1 ds e^{-sX} (\partial_t X) e^{sX} $.

\subsubsection{Resonant rotating Zeeman field}
The Hamiltonian in Eq.~\eqref{eq:H_d} holds generally. 
However, to clearly illustrate the effect of the small additional driving in Eq.~\eqref{eq:TM-pos}, we now focus for concreteness on the rotating Zeeman field $\pmb{\omega}_B^\text{P}$ defined in Eq.~\eqref{eq:IZ-rot}.
In this case, the Hamiltonian $H_\text{TM}$ simplifies as
\begin{subequations}
\label{eq:HTM}
\begin{align}
\frac{H_\text{TM}}{\hbar}&=\frac{\Delta}{2}\sigma_3- \left[\omega_\lambda e^{-i \omega_d t}-i\frac{\Omega}{2}\left(1-e^{-2i \omega_d t}\right)\right] \sigma_+ +\text{h.c.} \\
&\approx \frac{\Delta}{2}\sigma_3- \frac{\Omega}{2}\sigma_2 \ .
\end{align}
\end{subequations}
We defined here the Rabi frequency $\Omega=\omega_d Z/\lambda\ll \omega_d$, and we note that, in the second line, we used the conventional rotating wave approximation and neglected terms rotating at the fast frequency $\omega_d \gg \Delta, \omega_\lambda, \Omega$. We also introduced  $\sigma_\pm=\sigma_1\pm i\sigma_2$ and $\text{h.c.}$ indicates the hermitian conjugate.

The Rabi frequency $\Omega$ induces an additional rotation of the moving spin. 
At resonance $\Delta=0$, the spin dynamics in the rotating frame is captured by the unitary time evolution $U_\Omega= e^{i \Omega t\sigma_2 /2 } $, and thus in the original frame 
\begin{equation}
\label{eq:UTM}
U_\text{TM}=e^{-i\frac{\theta_B[\bar{z}(t)]\textbf{n}_B[\bar{z}(t)]\cdot\pmb{\sigma}}{2}}e^{-i\frac{\omega_d t\sigma_3}{2}}e^{i \frac{\Omega t\sigma_2 }{2} }e^{i\frac{\theta_B(0)\textbf{n}_B(0)\cdot\pmb{\sigma}}{2}} \ .
\end{equation}
The time evolution of the spin expectation values obtained starting from a spin state originally in the groundstate $|\downarrow\rangle$ are provided in Fig.~\ref{fig:filter_functions_dynamical}(a).
Even a small driving term $Z\ll l$ produces non-trivial spin dynamics, as we observe by comparing the solid and dashed curves, that correspond to the case $Z=0.01 l$ and $Z=0$, respectively. The spin dynamics  in the resonant case presents fast oscillations with frequency $\omega_d$ weighted by envelopes oscillating at the smaller frequencies $\Omega$ and $\omega_\lambda$.

This non-trivial deterministic spin dynamics also strongly modifies the response of the qubit to noise.
First, with a finite the Rabi driving, the dominant longitudinal component of the filter function is aligned to the  $\textbf{n}_2$ direction, thus leading to  
\begin{equation}
\label{eq:F33_D}
F= \frac{\nu}{N}\int_{0}^{t} d\tau d\tau' e^{-\frac{[\bar{z}(\tau)-\bar{z}(\tau')]^2}{2l^2}}  e^{i\omega(\tau'-\tau)}
[\hat{R}_{Z}^T(\tau)\hat{R}_{Z}(\tau')]_{22} \ .
\end{equation}
In contrast to Eq.~\eqref{eq:F33}, the kernel of the integral depends on $\hat{R}_{Z}(t) \textbf{n}_{2}= \hat{R}_1[2\bar{z}(t)/\lambda]\hat{R}_3[\omega_d t]\textbf{n}_{2}$, and oscillates at the high frequency $\omega_d=\omega_B$. 
For this reason, one might expect $F$ to be peaked at high frequencies.
However, we emphasize that the wavefunction contribution $ e^{-{[\bar{z}(\tau)-\bar{z}(\tau')]^2}/{2l^2}}$ also oscillates at frequency $\omega_d$, because $\bar{z}(t)$ contains the rapidly oscillating term $\propto Z$, see Eq.~\eqref{eq:TM-pos}, and thus $F$ has finite weight also at low frequency, where the noise is the largest.

The exact filter function obtained by integrating  Eq.\eqref{eq:F33_D} is shown in Fig.~\ref{fig:filter_functions_dynamical}(b) with a solid red line. The integral can be performed analytical for small values of $Z$, but the results are lengthy and we do not report them here. However, we note that by focusing on the  dominant low-frequency terms, $F$ is well approximated by   
\begin{equation}
\label{eq:F33_D_app}
F\approx\frac{Z^2}{l^2 }\frac{\nu}{N}\frac{t}{8 \omega_l }\frac{\omega^2}{\omega_l^2}\left[f_G\!\left(\frac{\omega-2\omega_\lambda}{\omega_l}\right) +f_G\!\left(\frac{\omega+2\omega_\lambda}{\omega_l}\right) \right] \ ,
\end{equation}
see the dashed orange line in Fig.~\ref{fig:filter_functions_dynamical}(b).
Importantly, because of the resonant dynamical decoupling, the low-frequency noise is efficiently filtered out by the additional polynomial factor $\omega^2$ in the filter function.

The polynomial factor $\omega^2$ in $F$ yields the exponential decay of the average shuttling fidelity
\begin{subequations}
\label{eq:fid_TM}
\begin{align}
\bar{\mathcal{F}}_\text{TM}&=e^{-t/T_\varphi^\text{TM}} \ , \\
 T_\varphi^\text{TM}&\approx\frac{4l^2}{\eta  Z^2} T_\varphi^L \left[1+ \sqrt{2 \pi }\frac{ l }{ \lambda }e^{\frac{2 l^2}{\lambda ^2}} \text{erf}\left(\frac{\sqrt{2} l}{\lambda }\right)\right]^{-1} \ .
\end{align}
\end{subequations}
Compared to the case with  $Z=0$ where the time scale is $T_\varphi^L$ in Eq.~\eqref{eq:inf-INhomo_g}, $T_\varphi^\text{TM}$ is substantially enhanced by the large factor $l^2/\eta Z^2\gg 1 $.
The dependence of time constant $ T_\varphi^\text{TM}$ on the inhomogeneity of the field $l/\lambda$ is illustrated in the inset of Fig.~\ref{fig:filter_functions_dynamical}(b).
Strikingly, the decay time $T_\varphi^\text{TM}$ is significantly larger than $T_\varphi^L$ when the Zeeman field is not strongly inhomogeneous $l/\lambda\lesssim 1$, but it becomes smaller at $l/\lambda\gtrsim 1$. 

The enhancement in average shuttling fidelity induced by the time-modulation of the position can be clearly observed in Fig.~\ref{fig:filter_functions_dynamical}(c) by comparing black and blue curves. 
At small values of $l/\lambda$ (solid lines), there is a substantial improvement in the coherence of the shuttling process that is due to the resonant dynamical decoupling induced by $Z$. In contrast to the $Z=0$ case, where $l/\lambda\gtrsim 1$ is required to filter out low frequency noise (dashed lines), the time-modulation enables a high shuttling fidelity also in the regime where the Zeeman energy is weakly renormalized by the factor $e^{-l^2/\lambda^2}$. 

We also note that the high-frequency components of  $F$ in Eq.~\eqref{eq:F33_D} produce the additional high-frequency terms
\begin{equation}
F_\text{HF}\approx \frac{\nu}{N}\frac{t }{2\omega_l }\left[f_G\!\left(\frac{\omega-\omega_B}{\omega_l}\right) +f_G\!\left(\frac{\omega+\omega_B}{\omega_l}\right) \right] \ ,
\end{equation}
whose functional form resembles Eq.~\eqref{eq:F_FR,I}, but with shifted frequency $2\omega_\lambda\to \omega_B$.
These corrections modify the fidelity as $\bar{\mathcal{F}}_\text{TM}\to \bar{\mathcal{F}}_\text{TM} e^{-t/T_\varphi^B}$, with time constant $T_\varphi^B= T_\varphi^L e^{-2l^2/\lambda^2+ \omega_B^2/\omega_l^2}\gg T_\varphi^\text{TM} $ for small values of $\omega_l/\omega_B\ll 1$.

\subsubsection{\label{sec:PD}Finite detuning and phase driving}
An homogeneous detuning $\Delta$ in $H_\text{TM}$ in Eq.~\eqref{eq:HTM} tilts the Rabi rotation by an angle $\varphi=\arctan(\Delta/\Omega)$ around the $\textbf{n}_1$-axis and speeds up the Rabi frequency by $\Omega\to \sqrt{\Omega^2+\Delta^2}$. 
The detuning causes incomplete Rabi oscillations with probability $P=\Omega^2/(\Omega^2+\Delta^2)$ and the typical Rabi chevron pattern measured in Rabi experiments.
Assuming a large driving field $\Omega$ compared to $\Delta$, the angle $\varphi\approx \Delta/\Omega \ll 1$ causes the appearance of a competing decay time for the average shuttling fidelity in Eq.~\eqref{eq:fid_TM} that modifies as $\bar{\mathcal{F}}_\text{TM}\to \bar{\mathcal{F}}_\text{TM} e^{-t/T_\varphi^\Delta}$, with
\begin{equation}
\label{eq:homo-det}
T_\varphi^\Delta\approx T_\varphi^L/ \varphi^2 +\mathcal{O}(\varphi^3) \ ,
\end{equation}
where the decay time $T_\varphi^L$ of adiabatic shuttling is given in Eq.~\eqref{eq:inf-INhomo_g}.

First, comparing $T_\varphi^\Delta$ to $T_\varphi^\text{TM}$, we find that  $T_\varphi^\Delta$ dominates when the Zeeman field is largely inhomogeneous $l/\lambda\gtrsim 1$ and when the power spectrum $S(\omega)$ of the noise strongly deviates from the $1/|\omega|$ trend, i.e. at large values of $\eta$ in Eq.~\eqref{eq:S_1/f}. 
However, even in this case, we emphasize that sufficiently close to resonance [$\varphi\ll 1$] $T_\varphi^\Delta\gg T_\varphi^L$ thus showing that time-modulation provides a substantial advantage compared to adiabatic driving. 

We also point out that an inhomogeneous detuning $\Delta(\bar{z})$, which can originate in experiments from local modulations of the $g$ factor or the magnetic fields, in general only results in an additional a small correction to the fidelity in Eq.~\eqref{eq:homo-det}. 
In particular, we focus here on the following detuning
\begin{equation}
\label{eq:Delta_phase}
\Delta[\bar{z}(t)]= \Delta_0+ \Delta_1 \cos\left[\frac{2\bar{z}(t)}{\lambda_\Delta}\right]\approx\Delta_0+  \Delta_1 \cos\left(2\omega_\Delta t\right) \ ,
\end{equation}
where we introduced $\omega_\Delta=\bar{v}/\lambda_\Delta$ and $\Delta_0=\omega_B-\omega_d$, with $\omega_B$ being the average Zeeman energy during shuttling.

With this inhomogeneous detuning, the Hamiltonian $H_\text{TM}$ in Eq.~\eqref{eq:HTM} is modified to the phase driving Hamiltonian~\cite{bosco2023phase}
\begin{equation}
\label{eq:PD-H}
{H_\text{PD}}\approx \frac{\hbar\omega_B}{2}\sigma_3+\frac{\hbar\Delta_1}{2}\cos\left(2\omega_\Delta t\right)\sigma_3- \hbar\Omega \sin(\omega_d t)\sigma_1 \ ,
\end{equation}
where the driving field has two tones and couplea to the transversal (Rabi driving $\propto \Omega \sigma_1$) and longitudinal (phase driving $\propto \Delta_1 \sigma_3$) spin degrees of freedom.
For clarity, here we report the Hamiltonian before performing the rotating frame transformation $e^{-i\omega_d t \sigma_3/2}$ of $U_Z^\text{TM}$, i.e. without the rotation $ \hat{R}_3(\omega_d t)$ in $H_\text{TM}$, see Eq.~\eqref{eq:H_d}.

As demonstrated in Ref.~\cite{bosco2023phase}, in general cases only off-resonant phase driving, with frequency $\omega_\Delta \sim \Omega$, significantly impacts the spin dynamics. For this reason, in Eq.~\eqref{eq:Delta_phase}, we discarded fast rotating phase driving terms oscillating at frequencies $\omega_d$.
In contrast, Rabi driving only impacts the spin dynamics when close to resonance $\omega_d\sim \omega_B$, and for this reason we neglect slowly rotating Rabi driving terms oscillating at frequencies $\omega_l$.

For small values of the modulation $\Delta_1\lesssim\omega_\Delta$, the effect of phase driving is negligible and one can safely operate at $\Delta_0=0$, i.e., by using a microwave pulse resonant with the average of the inhomogeneous Zeeman energy $\omega_B$.
For larger values of $\Delta_1\gtrsim\omega_\Delta$, phase driving introduces additional interesting dynamics in the spin evolution~\cite{bosco2023phase}. First, operating at a finite $\Delta_0$, enables additional resonant dynamics of the spin at $\Delta_0=\pm 2 m \omega\Delta$, with integer $m$, where the Rabi frequency is rescaled by $\Omega J_m(\Delta_1/\omega_\Delta)$.  Here $J_m(x)$ is the $m^\text{th}$ Bessel function. This additional resonant dynamics will also effectively filter out low-frequency noise.
Moreover, as discussed in Ref.~\cite{bosco2023phase},  even for small values of $\Delta_1\lesssim \omega_\Delta$, by fine tuning the Rabi frequency to $\Omega\sim 2\omega_\Delta $, we expect that additional resonant dynamics could substantially enhance the filtering out of dominant noise sources, further improving the average shuttling fidelity.

\subsubsection{Precessing Zeeman field}
Finally, we discuss the role of a precessing Zeeman field when the position is time-modulated.
These effects can be nicely described by our theory and in particular by $H_\text{TM}$  in Eq.~\eqref{eq:H_d}. 
By considering for concreteness  the precessing Zeeman field $\pmb{\omega}_B^\text{N}$ in  Eq.~\eqref{eq:SNutation-Zeeman} which enables spin nutation, and using Eq.~\eqref{eq:theta_B^d}, we find that  for small values of $A$, the driving term in $H_\text{TM}$ modifies as
\begin{equation}
\delta\pmb{\theta}_B=\frac{2}{\lambda} \left[\textbf{n}_1 - A\sin \left(\frac{4 \bar{z}}{\lambda }\right)\textbf{n}_2+ A\cos \left(\frac{4 \bar{z}}{\lambda }\right)\textbf{n}_3\right] + \mathcal{O}\left(A^2\right) \ ,
\end{equation}
where we only kept the terms to linear order in $A$, and we restrict ourselves to the analysis of the case $\lambda=\lambda_N$.

In this case, there are two leading corrections to the spin dynamics. 
In particular, we note that the last contribution in the expansion gives rise to a phase driving, see Eq.~\eqref{eq:PD-H}, with frequency $4\omega_\lambda$ and amplitude $2\omega_\lambda A$. As argued in Sec.~\ref{sec:PD}, for small values of $A$ and far from the fine-tuned resonance condition $4\omega_\lambda\sim \Omega$, this term has negligible effect. 

Moreover, the transversal term comprises a far detuned pulse with frequency $4\omega_\lambda$ that does not significantly contribute  to the spin dynamics and the frequency-modulated nearly-resonant term $-\omega_d AZ\sin(\omega_d t)\sin(4 \omega_\lambda t) $. In the RWA, this term yields an additional  transversal  Rabi driving. 
When off-resonant and $\Omega\gg 4 \omega_\lambda$, this term is negligible and thus we do not explore it further in this work.
Interestingly, however, we envision that this frequency-modulated driving could provide an additional effective filtering  of the noise, that is analogous to the frequency modulation in SMART dressed qubit protocols in global fields~\cite{Laucht2017,PhysRevB.104.235411,doi:10.1063/5.0096467}.  An optimized pulse-shaping  could also further enhance the fidelity \cite{Yang2019_1,PhysRevA.100.022337,PhysRevApplied.18.054090,rimbach2023simple}.

\subsection{Fast shuttling in weakly inhomogeneous fields}
In Sec.~\ref{sub: Examples}, we showed that fully rotating Zeeman field enables an effective way to intrinsically dynamically decouple a shuttled spin from low-frequency noise, thus resulting in a high shuttling fidelity. 
In particular, we focused on particles moving adiabatically with a constant velocity $\bar{v}\ll |\pmb{\omega}_B| l \ll \omega_o l$, which is small compared to both Zeeman and orbital energy. 
We also demonstrated in Sec.~\ref{sec:FMP} that the shuttling fidelity can be further improved by adding a small time-dependent modulation which is non-adiabatic with respect to $|\pmb{\omega}_B|$, but still adiabatic compared to  $\omega_o$.
Here, we show that a substantial  improvement in fidelity also occurs for incomplete rotations of the Zeeman field, when the constant shuttling velocity is non-adiabatic compared to the Zeeman field $|\pmb{\omega}_B|$, but remains adiabatic compared to  $\omega_o$.

For concreteness, we consider the weakly inhomogeneous Zeeman field 
\begin{equation}
\label{eq:Zeeman-DD}
\pmb{\omega}_B^D(\bar{z})= \omega_B \left[A \cos\left(\frac{2 \bar{z}}{\lambda}\right)\textbf{n}_1+ A \sin\left(\frac{2 \bar{z}}{\lambda}\right)\textbf{n}_2 +\textbf{n}_3 \right] \ ,
\end{equation} 
with $A\ll 1$ and a constant velocity motion with $\bar{z}=\bar{v} t$.
This Hamiltonian accurately describes a residual homogeneous magnetic field in electronic systems with nanomagnets~\cite{10.1063/5.0139670} and hole heterostructures, for example in planar germanium, presenting an incomplete tilting of the $g$-tensor~\cite{van2023coherent}. 

When $\bar{v}$ is adiabatic compared to the Zeeman field $\bar{v}\ll \omega_B l$, the shuttling fidelity is dominated by the dephasing accumulated by the homogeneous component of the Zeeman field aligned along $\textbf{n}_3$, see Sec.~\ref{sub: Examples}.
However, here we focus instead on a different case, where $ \bar{v}\sim \omega_B l$ and show that in this case there are resonant conditions for $\bar{v}$ that can substantially filter out low-frequency noise, still providing a large enhancement in the shuttling fidelity. 
We note that, as derived in Appendix~\ref{App:Non_adiabatic}, in this case corrections to Eq.~\eqref{eq:H_IZ} are $ \propto A \omega_\lambda\omega_B/\omega_o$, with $\omega_\lambda=\bar{v}/\lambda$ and remain negligible compared to the leading terms $\propto A\omega_B$ also in this case. 

The resonance condition in this case is straightforwardly recognizable by moving to a rotating frame with frequency $2\omega_\lambda$ by the transformation $e^{-i\omega_\lambda t\sigma_3}$. In this frame, we immediately recognize the time-independent Rabi Hamiltonian
\begin{equation}
\label{eq:H_D}
H_\text{D}=\frac{\hbar\Delta}{2}\sigma_3+ \frac{\hbar\Omega}{2}\sigma_1 \ , \ \Omega=A\omega_B  \ , \ \Delta=\omega_B-2\omega_\lambda \ ,
\end{equation}
describing Rabi oscillation of the spin with Rabi frequency $\Omega$ at the resonance $\Delta=0$, see also Eq.~\eqref{eq:HTM}.

We  focus now on local noise sources and we can straightforwardly verify that the longitudinal component of the filter function $F_\text{D}^L$ is equivalent to $F_\text{P}^L$  that is given in Eq.~\eqref{eq:F_FR,I}. 
We then find the shuttling fidelity  
\begin{equation}
  \bar{\mathcal{F}}_\text{D}^L\approx e^{-{t}/{T_\varphi^\text{D}}} \ , \ 
  T_\varphi^\text{D}= \omega_l \left(\sqrt{\frac{\nu}{N}} T_\varphi\right)^2 e^{\frac{2l^2}{\lambda^2}} \ .
\end{equation}
This result is equivalent to Eq.~\eqref{eq:inf-INhomo_g}, however, we emphasize that because in this case the Zeeman field is not fully rotating, the Zeeman energy $\omega_B$ is not rescaled by the small prefactor $e^{-l^2/\lambda^2}$. As a result at large values of  the ratio $l/\lambda\gtrsim 1$ the Zeeman energy remains large, while the fidelity is rapidly improved. This critical difference between this approach and the one in Sec.~\ref{sub: Examples} is clearly illustrated by comparing the dashed and solid gray lines in Fig.~\ref{fig:filter_functions_SOI}(b), corresponding to the Zeeman fields in the two situations, which yield the same shuttling fidelity (blue line). 

We stress that the condition $\Delta=0$ is within reach of current experiments.
Considering $\omega_B/2\pi=1$~GHz and $\pi\lambda=50$~nm, which corresponds to either the spacing between neighbouring nanomagnets for electronic systems or the gate spacing determining the tilt of the g-factor in planar hole heterostructures, we find that high-fidelity shuttling can be achieved at feasible velocities $\bar{v}=\omega_B \lambda/2=50$~m/s~\cite{PRXQuantum.4.020305}. Our protocol thus enables at the same time high-fidelity and fast shuttling even in the presence of residual large homogeneous Zeeman fields.

\section{\label{sec:Conculsion} Conclusion}
In this work, we showed that the fidelity of the spin shuttling can be substantially enhanced  by engineering highly inhomogeneous Zeeman fields. 
We related this surprising effect to the non-trivial deterministic dynamics of the spin during its motion, which filters out the dominant low-frequency components of the noise.
This intrinsic dynamical decoupling of low-frequency noise is a general feature that appears in a wide variety of relevant experimental cases, including hole nanostructures in silicon and germanium as well as in electronic systems with artificial spin-orbit fields induced by micromagnets.  
We propose a framework to describe many scenarios where spins are shuttled in an inhomogeneous Zeeman field caused by rotation of principal axes of $g$-tensors, inhomogeneous magnetic field, and SOI.
We also include a detailed analysis of different sources of noise, that affect the shuttled spin in a global or local way. Despite some qualitative and quantitative differences in these cases, we confirm that an inhomogeneous Zeeman field  improves shuttling fidelity independent of the noise locality.
We also propose protocols where the spin is moved non-adiabatically compared to the Zeeman energy, that enable further dynamical decoupling of low-frequency noise and thus can significantly improve the coherence of shuttling.
Our findings clearly demonstrate that highly efficient shuttling can be reached in materials with large SOI and inhomogeneous Zeeman fields, and that these systems are not only ideal hosts for compact spin-qubit architectures, but also for long-range spin qubit connectivity, and are thus ideal candidates for future large-scale quantum computers.

\acknowledgements{
We thank Andreas Fuhrer, Michele Aldeghi, Andras Palyi, Bence Hetényi, and Maria Spethmann for useful discussions.
This work was supported by the Swiss National Science Foundation, NCCR SPIN (grant number 51NF40-180604), and  the Georg H. Endress Foundation. 
}

\appendix
\section{Non-adiabatic corrections}
\label{App:Non_adiabatic}
We now discuss in more detail the condition of adiabaticity of the quantum dot motion compared to the orbital degrees of freedom.
We stress again that in the main text, we occasionally lift the condition of adiabaticity compared to the Zeeman field $\omega_B/2\pi \lesssim 10$~GHz, but the shuttling is always adiabatic compared to the orbital splitting $\omega_o/2\pi\gtrsim 1$~THz.
First, we derive with a simple perturbative treatment the expected corrections to the model presented in the main text, and then we verify these corrections by showing that they match an exactly solvable simple case.

\subsection{Perturbative treatment}
Our derivations in the main text always assume that the quantum dot motion remains adiabatic compared to the orbital degree of freedom. We now discuss the validity of this approximation by using a simple model, that includes perturbatively the contribution of the next excited orbital state.

In particular, we now include in our derivation of Eq.~\eqref{eq:H_IZ} the effect of the neglected dynamical term $-i p \partial_t \bar{z} $, originating from the time-dependence of the state $\psi[z-\bar{z}(t)]$. 
The expectation value of this term in the ground state vanishes. However, this term provides a coupling to the first excited state $\psi_1[z-\bar{z}(t)]$; assuming a harmonic potential $\psi_1(z)=H_1(z/l) e^{-z^2/2l^2}/\pi^{1/4}\sqrt{2 l} $ with $H_1$ being the first Hermite polynomial. 

The  effective Hamiltonian acting on these two states is
\begin{equation}
H= \hbar\left(
\begin{array}{cc}
\frac{\pmb{\omega}_B(\bar{z})\cdot \pmb{\sigma} }{2}& -\frac{l\partial_{\bar{z}}\pmb{\omega}_B(\bar{z})\cdot \pmb{\sigma} }{2\sqrt{2}} + \frac{\partial_{t}\bar{z} \sigma_0}{\sqrt{2}l} \\
-\frac{l\partial_{\bar{z}}\pmb{\omega}_B(\bar{z})\cdot \pmb{\sigma} }{2\sqrt{2}}  +  \frac{\partial_{t}\bar{z} \sigma_0}{\sqrt{2}l} & \frac{\pmb{\omega}_B^{1}(\bar{z})\cdot \pmb{\sigma}}{2} + \omega_o \sigma_0
\end{array} 
\right)  \ ,
\end{equation}
where  we introduce $\pmb{\omega}_B^{1}(\bar{z})=\int dz |\psi_1(z)|^2\tilde{\pmb{\omega}}_B(z+\bar{z}) $. We also use the relation   $\int dz \psi(z)\psi_1(z)\tilde{\pmb{\omega}}_B(z+\bar{z})=-l\partial_{\bar{z}}\pmb{\omega}_B(\bar{z})/\sqrt{2} $, valid for harmonic potential eigenfunctions and straightforward to derive using the Rodrigues formula defining the Hermite polynomials.

By using second order perturbation theory, we find the effective Hamiltonian for the ground state 
\begin{equation}
H_\text{}=\frac{\hbar }{2}\left[\pmb{\omega}_B(\bar{z})+\frac{\partial_t \bar{z}}{\omega_o}\partial_{\bar{z}}\pmb{\omega}_B(\bar{z})\right]\cdot\pmb{\sigma} \ . \
\end{equation}
The corrections arising from the orbital non-adiabaticity of the motion scale with $\sim \omega_\lambda\omega_B/\omega_o $.
In our work, these corrections are most significant when we lift the Zeeman field adiabaticity  condition, in which case $\sim \omega_B^2/\omega_o $, and they still produce small terms that are quadratic in the magnetic field.

\subsection{\label{sec:Exact}Exact solution with SOI}
Here, we validate the perturbative results just derived by presenting an exact solution for the time-dependent Schrodinger equation, which fully accounts for non-adiabatic corrections.
This solution describes a spin confined in a quantum dot moving in a homogeneous Zeeman aligned to a possibly time-dependent SOI field with a fixed direction. 

We consider the following one-dimensional Hamiltonian
\begin{equation}
H=\frac{p^2}{2m}+\frac{m\omega_o^2}{2}z^2+ v(t) p \sigma_3+ m\omega_o^2 \bar{z}(t) z+ \frac{\hbar \omega_B }{2} \sigma_3\ ,
\end{equation}
where SOI and Zeeman fields are aligned to the $\textbf{n}_3$ direction.
Introducing the usual orbital bosonic ladder operators $a$ and $ a^\dagger$, the harmonic length $l$, the time-dependent spin-orbit length $\lambda_{s}(t)=\hbar/m v(t)$, we can rewrite this Hamiltonian as
\begin{equation}
\frac{H}{\hbar\omega_o}= a^\dagger a+\frac{\omega_B}{2\omega_o} \sigma_3+ \frac{il}{\sqrt{2}\lambda_{s}(t)} (a^\dagger-a) \sigma_z+ \frac{\bar{z}(t)}{\sqrt{2}l} (a^\dagger+a)  \ .
\end{equation}
We move to a spin-dependent rotating frame by the unitary operator
\begin{equation}
U_E(t)=e^{-i t(\omega_o a^\dagger a+\omega_B \sigma_3/2)} \ ,
\end{equation}
 yielding
\begin{subequations}
\begin{align}
\frac{H_R}{\hbar\omega_o}&= \alpha(t)a^\dagger+\alpha^\dagger(t) a  \ , \\
 \alpha(t)&=\frac{e^{i\omega_o t}}{\sqrt{2}}\left[\frac{\bar{z}(t)}{l}+\frac{il}{\lambda_{s}(t)}\sigma_3 \right]
\ ,
\end{align}
\end{subequations}
where we used $U_E^\dagger(t) a U_E(t)=ae^{-i\omega_o t}$. 

The time-evolution operator of the system can then be formally found as 
\begin{equation}
U(t)=U_E(t)\mathcal{T}e^{-i\int_0^{t}H_R(\tau)d\tau/\hbar}  \ .
\end{equation}
In our case, this equation can be evaluated exactly because the spin sector remains diagonal during the time evolution and the problem is quadratic in the orbital degree of freedom.
The explicit exact solution of time-ordered exponential is obtained by a second-order Magnus expansion~\cite{ZEUCH2020168327}: because $[a,a^\dagger]=1$ and higher order commutators coming from the expansions vanish and the result of the second-order expansion is exact.

We thus obtain 
\begin{equation}
U(t)=U_E(t)e^{-i\phi(t)}D\left[\Gamma(t) \right]  \ .
\end{equation} 
We introduced the conventional quantum optical displacement operator $D(x)=e^{x a^\dagger-x^\dagger a }$, and the spin-dependent phase-space shift $\Gamma(t)$ and  phase $\phi(t)$ are
\begin{subequations}
\begin{align}
\Gamma(t)&=-i \omega_o \int_0^t \alpha(\tau)d\tau  \ , \\
\phi(t)&=i\omega_o^2 \int_0^td\tau\int_0^\tau d\tau' \frac{\alpha(\tau)\alpha^\dagger(\tau')-\alpha^\dagger(\tau)\alpha(\tau')}{2} \ .
\end{align}
\end{subequations}

As a concrete example, we consider the case $\bar{z}(t)= \bar{v} t$ and a time-independent $\lambda_{s}$, in which case
\begin{subequations}
\begin{align}
\Gamma(t)&=-\frac{e^{i \omega _o t}}{\sqrt{2}}\left[\theta_1(t) +i \frac{l }{\lambda_{s}}\left(1-e^{-i \omega _o t}\right) \sigma_3  \right]\ , \\
\theta_1(t)&=\omega_l t +i \frac{\omega_l}{\omega_o}\left(1-e^{-i \omega _o t}\right)\ , \\
\phi(t)&=\phi_0-\left[\frac{\omega_s t}{2}\left[1+\cos \left(\omega _o t\right)\right]- \frac{\omega_s}{\omega_o} \sin \left(\omega_o t\right)\right]  \sigma_3\ ,
\end{align}
\end{subequations}
where $\phi_0$ is a trivial global spin-independent phase, and $\omega_{l}=\bar{v}/l$ and $\omega_{s}=\bar{v}/\lambda_s$ as in the main text.

We focus on the time-evolution of the orbital ground states of $H_R$ at time $t=0$ and centred at $\bar{z}(0)=0$:
\begin{equation}
|\psi_{\uparrow\downarrow}\rangle_0=D\left(-\frac{i l \sigma_3}{\sqrt{2} \lambda_{s}}\right) |0, \tilde{\uparrow}\tilde{\downarrow} \rangle  \ ,
\end{equation}
where the $\tilde{\uparrow}\tilde{\downarrow}$ spins are the pseudo spin degrees of freedom defined by removing the SOI by the usual transformation  $D\left(-{i l \sigma_3}/{\sqrt{2} \lambda_{s}}\right)$.
Time evolved state at time $t$ is 
\begin{multline}
|{\psi}_{\uparrow\downarrow}\rangle(t)=e^{-i\frac{\omega_B t}{2}\sigma_3-i\phi(t)} e^{i\frac{l }{\sqrt{2} \lambda_{s}}\text{Re}[\Gamma(t)]\sigma_3}\\
\times D\left[\left(\Gamma(t) -\frac{i l}{\sqrt{2} \lambda_{s}}\sigma_3\right)e^{-i\omega_o t}\right] |0, \tilde{\uparrow}\tilde{\downarrow} \rangle \ .
\end{multline}
In particular, when $\bar{z}(t)=\bar{v} t$, we find
\begin{subequations}
\begin{align}
|{\psi}_{\uparrow\downarrow}(t)\rangle & =e^{-i\frac{\omega_B t+\theta_0(t)}{2} \sigma_3} D\left[-\frac{\theta_1(t)}{\sqrt{2}}-i \frac{l }{\sqrt{2}\lambda_{s}} \sigma_3\right] |0,\tilde{\uparrow}\tilde{\downarrow}\rangle \\
&=e^{-i\frac{\omega_B t+2\theta_0(t)}{2} \sigma_3}D\left[-\frac{\theta_1(t)}{\sqrt{2}}\right]  D\left[-\frac{i l \sigma_3}{\sqrt{2}\lambda_{s}} \right]|0, \tilde{\uparrow}\tilde{\downarrow} \rangle \\
&=e^{-i\frac{\omega_B t+2\theta_0(t)}{2} \sigma_3}D\left[-\frac{\theta_1(t)}{\sqrt{2}}\right] |{\psi}_{\uparrow\downarrow}\rangle_0
 \ ,
\end{align}
\end{subequations}
where we introduce the spin-dependent phase shift
\begin{equation}
\theta_0(t)=-{\omega_s t}+\frac{\omega_s}{\omega _o }\sin\left(\omega _o t\right) \ .
\end{equation}
In this simple case, it is clear that the non-adiabatic corrections provide fast oscillations of the angles $\theta_{0,1}(t)$ that become suppressed as $\omega_l\sim \omega_s\ll \omega_o$. 

More generally, by averaging out the fast oscillations of period $1/\omega_o$, one can generalize these results as
\begin{equation}
\theta_0(t)\approx - \bar{z}(t)/\lambda_{s}  \ , \ \ \text{and} \   \ \theta_1(t)\approx \bar{z}(t)/l \ .
\end{equation}
As expected, we note that the non adiabatic corrections are $\propto \omega_{l,s}/\omega_o$ and result in additional oscillations terms $\propto e^{-i \omega_o t}$ that we neglect in our adiabatic approximation.

\section{\label{sec:Rot} Rotation matrices}
Here, we provide an explicit expression for the rotation matrices used in the main text.
The unitary operator
\begin{equation}
    U=e^{-i\theta \textbf{n}\cdot\pmb{\sigma}/2} \ ,
\end{equation}
with unit vector $\textbf{n}=(n_1,n_2,n_3)$ (such that $\textbf{n}\cdot\textbf{n}=1$), transforms a vector of Pauli matrices $\pmb{\sigma}=(\sigma_1,\sigma_2,\sigma_3)$ as
\begin{equation}
 U^\dagger \pmb{\sigma} U=\hat{R}_\textbf{n}(\theta)\pmb{\sigma} \ .
\end{equation}
The counterclockwise rotation matrix $\hat{R}_{\textbf{n}}(\theta)$ rotates a vector by an angle $\theta$ around $\textbf{n}$  and is given by
\begin{widetext}
\begin{equation}
\label{eq:general_rot}
    \hat{R}_{\textbf{n}}(\theta)=\left(
\begin{array}{ccc}
  \left(1-n_x^2\right)\cos (\theta )+n_x^2 & n_x n_y[1-\cos (\theta )]-\sin (\theta ) n_z & n_x n_z[1-\cos (\theta )]+\sin (\theta ) n_y \\
n_x n_y [1-\cos (\theta )] +\sin (\theta ) n_z &  \left(1-n_y^2\right) \cos (\theta )+n_y^2 & n_y n_z[1-\cos (\theta )]-\sin (\theta ) n_x\\
 n_x n_z[1-\cos (\theta )]-\sin (\theta ) n_y & n_y n_z[1-\cos (\theta )]+\sin (\theta ) n_x &  \left(1-n_z^2\right)\cos (\theta )+n_z^2 \\
\end{array}
\right) \ .
\end{equation}
\end{widetext}
For convenience, we also define the rotation matrices $\hat{R}_i(\theta)$ around the $i=(1,2,3)$ axis as
\begin{subequations}
\label{eq:rot_wire}
\begin{align}
\hat{R}_1(\theta)&=\left(
\begin{array}{ccc}
 1 &  0 & 0  \\
 0 & \cos(\theta)   &  -\sin(\theta) \\
 0 &  \sin(\theta) &  \cos(\theta)  \\
\end{array}
\right) \ , \\
\hat{R}_2(\theta)&=\left(
\begin{array}{ccc}
  \cos(\theta)  &  0 & \sin(\theta) \\
  0 & 1 & 0 \\
  -\sin(\theta) &  0 & \cos(\theta)
\end{array}
\right) \ , \\ 
\hat{R}_3(\theta)&=\left(
\begin{array}{ccc}
 \cos(\theta)   &  -\sin(\theta) & 0 \\
  \sin(\theta) &  \cos(\theta) & 0 \\
  0 &  0 & 1
\end{array}
\right) \ .
\end{align}
\end{subequations}
and we report the useful relation 
\begin{equation}
\label{eq:Rot_general_vect}
\hat{R}_\textbf{n}(\theta)\textbf{A}=\textbf{n}(\textbf{n}\cdot \textbf{A})+\cos\theta (\textbf{n}\times \textbf{A})\times \textbf{n}+\sin\theta (\textbf{n}\times\textbf{B}) \ .
\end{equation}

For the discussion in the main text, we are particularly interested in the solution of the equation
\begin{equation}
\hat{R}_{\textbf{n}}(\theta)\textbf{n}_3 = \left[\sin (\varphi ) \sin \left(\varphi _1\right),\sin \left(\varphi _1\right) \cos (\varphi ),\cos \left(\varphi _1\right)\right] \ ,
\end{equation}
which aligns a general vector parametrized by the angles $\varphi\in[0,2\pi)$ and $\varphi_1\in [0,\pi)$ to the $3$rd direction. Note that the rotation $\hat{R}_{\textbf{n}}(\theta)$ is straightforwardly decomposed as $\hat{R}_{\textbf{n}}(\theta)=\hat{R}_3(-\varphi)\hat{R}_1(-\varphi_1)$.
 A particular solution for the vector and angle of the combined rotations valid for $\cos(\varphi_1)\geq 0$ is
% \begin{subequations}
% \begin{align}
% \theta&=-\text{sgn}(\sin (\varphi )) \cos ^{-1}\left(\cos (\varphi ) \cos ^2\left(\frac{\varphi _1}{2}\right)-\sin ^2\left(\frac{\varphi _1}{2}\right)\right) \ , \\
% \textbf{n}&=\left[\cos \left(\frac{\varphi }{2}\right) \cos\left( \frac{\phi}{2} \right),-\sin \left(\frac{\varphi }{2}\right) \cos \left( \frac{\phi}{2} \right),\sin \left( \frac{\phi}{2} \right)\right] \ , \\
% \phi &= \cos ^{-1}\left(\frac{\cos ^2\left(\frac{\varphi }{2}\right)-\left(1+\sin ^2\left(\frac{\varphi }{2}\right)\right) \cos \left(\varphi _1\right)}{1+\sin ^2\left(\frac{\varphi _1}{2}\right)-\cos (\varphi ) \cos ^2\left(\frac{\varphi _1}{2}\right)}\right) \ .
% \end{align}
% \end{subequations}
 \begin{subequations}
 \begin{align}
 \theta&=-\text{sgn}(\sin\varphi ) \cos ^{-1}\!\left(\frac{\cos\varphi+(1+\cos \varphi ) \cos \varphi _1-1}{2} \right) \ , \\
 \textbf{n}&=\left[\cos \left(\frac{\varphi }{2}\right) \cos\left( \frac{\phi}{2} \right),-\sin \left(\frac{\varphi }{2}\right) \cos \left( \frac{\phi}{2} \right),\sin \left( \frac{\phi}{2} \right)\right] \ , \\
 \phi &=-2 \sin ^{-1}\left[ \tan\left(\frac{\varphi}{2}\right)\cot\left(\frac{\theta}{2}\right)\right] \ .
 \end{align}
 \end{subequations}

For small positive angles $\varphi_1$ around the third axis, one can Taylor expand this solution to the first order in $\varphi_1$, resulting in 
 \begin{subequations}
 \begin{align}
 \theta&=- \cos ^{-1}(\cos (\varphi )) \text{sgn}(\sin (\varphi ))  +\mathcal{O}(\varphi_1^2)  \ , \\
 \phi &= \pi-\frac{\varphi_1}{|\sin(\varphi/2)|}  +\mathcal{O}(\varphi_1^2)\ , \\
 \textbf{n}&=\left[\frac{\varphi _1 \text{sgn}\left(\sin \left(\frac{\varphi }{2}\right)\right)}{2}  \cot \left(\frac{\varphi }{2}\right)  ,\ \frac{-\varphi _1 \text{sgn}\left(\sin \left(\frac{\varphi }{2}\right)\right)}{2}  , \ 1 \right] \ , 
 \end{align}
 \end{subequations}
or equivalently, unwinding the phases,
 \begin{subequations}
 \begin{align}
 \theta&=- \varphi  \ , \\
 \textbf{n}&=\left[\frac{\varphi _1}{2} \cot\left(\frac{\varphi }{2}\right)   ,\ -\frac{\varphi _1}{2}  , \ 1 \right]  +\mathcal{O}(\varphi_1^2) \ ,
 \end{align}
 \end{subequations}
resulting in the vector
\begin{equation}
\hat{R}_{\textbf{n}}(\theta)\textbf{n}_3=\left[ \varphi_1 \sin(\varphi) , \  \varphi_1 \cos(\varphi) , \ 1\right] +\mathcal{O}(\varphi_1^2) \ .
\end{equation}

\section{\label{app:SOI-homo-inhomo} Intermediate-range noise sources}
We discuss in more detail the role of inhomogeneous noise with an intermediate range. We focus on systems with arbitrary SOI.
We consider to this aim the Hamiltonian $H_\text{1D}$ in Eq.~\eqref{eq:H_1d_SOI}, and the noise Hamiltonian
\begin{equation}
    \label{eq:H_N_inhomo-gen}
    H_N^z= \frac{1}{2n_0} \sum_k V(z-z_k)\pmb{h}_k(t)\cdot\pmb{\sigma} \ ,
\end{equation}
where the function $V(z-z_k)$ determines whether the noise sources are local [$V(z)=\delta(z)$] or global [$V(z)=n_0$]. We consider here an homogeneous Zeeman field, i.e., $ \tilde{\pmb{\omega}}_B({z})=\tilde{\pmb{\omega}}_B$.

We remove the SOI by the transformation $S$ in Eq.~\eqref{eq:trafo_S} and we project the total Hamiltonian onto the moving dot ground state wavefunction, resulting in the effective Hamiltonian
\begin{equation}
H= \frac{\hbar \pmb{\omega}_B(\bar{z})\cdot\pmb{\sigma}}{2}+\frac{1}{2} \tilde{\pmb{H}}(\bar{z},t) \cdot\pmb{\sigma} \ , 
\end{equation}
with [see Eq.~\eqref{eq:omegaB_SOI}]
\begin{widetext}
\begin{subequations}
\begin{align}
\pmb{\omega}_B(\bar{z})&=\tilde{\pmb{\omega}}_B \int_{-\infty}^{+\infty} d z |\psi(z-\bar{z})|^2  \hat{R}_{s}^T[2z/\lambda_s] \hat{R}_{ \delta(z)}^T[\phi_s(z)]    \   , \\
\tilde{\pmb{H}}(\bar{z},t)&=\frac{1}{n_0}\sum_k\pmb{h}_k(t)\int_{-\infty}^{+\infty}d z V(z-z_k)|\psi(z-\bar{z})|^2\hat{R}_{s}^T[2z/\lambda_s] \hat{R}_{ \delta(z)}^T[\phi_s(z)]\   .
\end{align}
\end{subequations}

The longitudinal component of the covariance matrix $\hat{\Sigma}_{33}$, which determines the average shuttling fidelity is 
\begin{equation}
\label{eq:covariance_inhomog}
    \hat{\Sigma}_{33}=\frac{1}{2\pi \hbar^2}\int_{-\infty}^{\infty} d\omega  \pmb{\eta}^\dagger(\omega, t) \hat{S}(\omega) \pmb{\eta}(\omega, t) \ ,
\end{equation}
with $\hat{S}_{ij}(\omega)=\int d t e^{i\omega t }\langle h_i(t)h_j(0)\rangle $ being a general anisotropic noise spectral function. We introduced the vector 
\begin{align}
\pmb{\eta}=\frac{1}{n_0} \int_{0}^{t} d\tau e^{-i\omega \tau}\sum_k\int_{-\infty}^{+\infty}d zV(z-z_k)|\psi(z-\bar{z}(\tau))|^2 \hat{R}_{s}^T[2z/\lambda_s] \hat{R}_{ \delta(z)}^T[\phi_s(z)]
\frac{\pmb{\omega}_B[\bar{z}(\tau)]}{|\pmb{\omega}_B[\bar{z}(\tau)]|} \  .
\end{align}
Assuming isotropic uncorrelated noise, $\hat{S}_{ij}(\omega)= \delta_{ij} S(\omega)$, we find that the longitudinal filter function is $\hat{F}_{33}=F = \pmb{\eta}^\dagger\cdot \pmb{\eta} $.

For global and local noise, we obtain, respectively,
\begin{equation}
\tilde{\pmb{H}}_\text{G}=\pmb{h} \int_{-\infty}^{+\infty} d z |\psi(z-\bar{z})|^2\hat{R}_{s}^T[2z/\lambda_s] \hat{R}_{ \delta(z)}^T[\phi_s(z)] \ ,   \ \  \text{and} \ \  \
\tilde{\pmb{H}}_\text{L}=\frac{1}{n_0}\sum_k|\psi(z_k-\bar{z})|^2\pmb{h}_k\hat{R}_{s}^T[2z_k/\lambda_s] \hat{R}_{ \delta(z_k)}^T[\phi_s(z_k)]  \ . 
\end{equation}
\end{widetext}
We define here $\pmb{h}=\sum_k\pmb{h}_k$. We emphasize that for global noise the SOI-induced rotation is independent of the location of the defects and affects noise and Zeeman field in the same way. Local noise, on the other hand, locally rotates the noise fluctuators yielding a qualitatively different effect compared to the Zeeman field.
 
This qualitative difference can be straightforwardly understood by considering a homogeneous SOI field, such as $\textbf{v}_\text{H}$ in Eq.~\eqref{eq:vh}. In this case $\phi_s(z)=0$ and we find 
 \begin{subequations}
 \begin{align}
\pmb{\omega}_B&=\tilde{\pmb{\omega}}_B^\parallel+ e^{-l^2/\lambda_s^2}\tilde{\pmb{\omega}}_B^\perp \hat{R}_{s}^T(2\bar{z}/\lambda_s)\ , \\
\tilde{\pmb{H}}_\text{G}&={\pmb{h}}^\parallel+ e^{-l^2/\lambda_s^2}{\pmb{h}}^\perp \hat{R}_{s}^T(2\bar{z}/\lambda_s) \ ,   \\
\tilde{\pmb{H}}_\text{L}&=\frac{1}{n_0}\sum_k|\psi(z_k-\bar{z})|^2\pmb{h}_k\hat{R}_{s}^T[2z_k/\lambda_s]  \ ,
\end{align}
\end{subequations}
where $\parallel$ and $\perp$ refer to components of the vectors parallel and perpendicular to the SOI $\textbf{n}_s$, respectively.

We now focus on the case where the Zeeman field is perpendicular to the SOI, e.g., $\textbf{n}_s= \textbf{n}_1$ and $\tilde{\pmb{\omega}}_B=\tilde{\omega}_B \textbf{n}_3$. In the interaction picture including the dynamics induced by the Zeeman field, the relevant longitudinal noise is
\begin{subequations}
\begin{align}
[\tilde{\pmb{H}}_\text{G}]_3&=e^{-l^2/\lambda_s^2}{\pmb{h}}\cdot\textbf{n}_3  \ ,   \\
[\tilde{\pmb{H}}_\text{L}]_3&=\frac{1}{n_0}\sum_k|\psi(z_k-\bar{z})|^2\pmb{h}_k \cdot\hat{R}_{s}^T[2(z_k-\bar{z})/\lambda_s] \textbf{n}_3  \ ,
\end{align}
\end{subequations}
resulting in [see Eqs.~\eqref{eq:f_FID}~and~\eqref{eq:F33I}]
\begin{align}
F_\text{G}=e^{-2l^2/\lambda_s^2}F_\text{FID}  \ , \ \ \text{and} \ \ F_\text{L}=F_\text{P}^L   \ .
\end{align}

\section{Exact results for the filter functions}
\label{sec:Filter_funct}
In this section, we report exact equations for filter function for a fully precessing Zeeman field and fidelities:
\begin{widetext} 
\begin{subequations}
\begin{align}
\label{eq:F_FR_exact}
  F_\text{P}& =\frac{2 \left(\omega _{\lambda }^2-\left(\omega _{\lambda }^2+\omega ^2\right) \cos (t \omega ) \cos \left(t \omega _{\lambda }\right)-2 \omega  \omega _{\lambda } \sin (t \omega ) \sin \left(t \omega _{\lambda }\right)+\omega ^2\right)}{\left(\omega ^2-\omega _{\lambda }^2\right)^2} \ , \\
\label{eq:Lorentzian_FR_decay}
\bar{\mathcal{F}}_\text{P}&= \exp\left[\frac{1}{8} i \pi  2^{\eta -1} \csc (\pi  \eta ) t^{2-\eta } T^{\eta -2} \left(-(t \omega_\lambda +i)^{\eta -1}+( t \omega_\lambda -i)^{\eta -1}+(-t \omega_\lambda -i)^{\eta -1}-(-t \omega_\lambda +i)^{\eta -1}\right) \right]\ , \\
 \label{eq:Lorentzian_FR_decay_in}
F_\text{P}^L&=\frac{1}{\omega_l^2}\text{Re}\left[ f\left(\frac{\omega-2\omega_\lambda}{\omega_l}, \omega_l t \right)\right] \ , \\
f(\omega, t)&=e^{-\frac{t^2+\omega ^2}{2}} \left[\sqrt{\frac{2}{\pi }} e^{\frac{\omega ^2}{2}} \left(\cos (t \omega )-e^{\frac{t^2}{2}}\right)+\frac{1}{2}e^{\frac{t^2}{2}} \left((t+i \omega ) \text{erf}\left(\frac{t+i \omega }{\sqrt{2}}\right)+(t-i \omega ) \text{erf}\left(\frac{t-i \omega }{\sqrt{2}}\right)+2 \omega  \text{erfi}\left(\frac{\omega }{\sqrt{2}}\right)\right)\right] \    .
    \end{align} 
 \end{subequations}
\end{widetext}

\bibliography{literature}

\end{document}